# Heavy Hitters and the Structure of Local Privacy

Mark Bun[*]    Jelani Nelson[†]    Uri Stemmer[‡]

November 13, 2017

**Abstract**

We present a new locally differentially private algorithm for the heavy hitters problem which achieves optimal worst-case error as a function of all standardly considered parameters. Prior work obtained error rates which depend optimally on the number of users, the size of the domain, and the privacy parameter, but depend sub-optimally on the failure probability.

We strengthen existing lower bounds on the error to incorporate the failure probability, and show that our new upper bound is tight with respect to this parameter as well. Our lower bound is based on a new understanding of the structure of locally private protocols. We further develop these ideas to obtain the following general results beyond heavy hitters.

- *Advanced Grouposition*: In the local model, group privacy for $k$ users degrades proportionally to $\approx \sqrt{k}$, instead of linearly in $k$ as in the central model. Stronger group privacy yields improved max-information guarantees, as well as stronger lower bounds (via "packing arguments"), over the central model.

- Building on a transformation of Bassily and Smith (STOC 2015), we give a generic transformation from any non-interactive approximate-private local protocol into a pure-private local protocol. Again in contrast with the central model, this shows that we cannot obtain more accurate algorithms by moving from pure to approximate local privacy.

## 1 Introduction

In the *heavy-hitters problem*, each of $n$ users holds as input an item from some domain $X$. Our goal is to identify all "heavy-hitters," which are the domain elements $x \in X$ for which many of the users have $x$ as their input. In this work we study the heavy-hitters problem in the *local model* of differential privacy (LDP), where the users randomize their data locally, and send differentially private reports to an untrusted server that aggregates them. In our case, this means that every user only sends the server a single randomized message that leaks very little information about the input of that user.

The heavy hitters problem is perhaps the most well-studied problem in local differential privacy [4, 17, 24, 3, 27]. In addition to the intrinsic interest of this problem, LDP algorithms for heavy-hitters provide important subroutines for solving many other problems, such as median estimation, convex optimization, and clustering [31, 26]. In practice, heavy-hitters algorithms under LDP have already been implemented to provide organizations with valuable information about their user bases while providing strong privacy guarantees to their users. Two prominent examples are by Google in the Chrome browser [12] and Apple in iOS-10 [33], making local heavy-hitter algorithms the most widespread industrial application of differential privacy to date.

---

[*]Department of Computer Science, Princeton University. mbun@cs.princeton.edu

[†]John A. Paulson School of Engineering and Applied Sciences, Harvard University. minilek@seas.harvard.edu. Supported by NSF grant IIS-1447471 and CAREER award CCF-1350670, ONR Young Investigator award N00014-15-1-2388 and DORECG award N00014-17-1-2127, an Alfred P. Sloan Research Fellowship, and a Google Faculty Research Award.

[‡]Dept. of Computer Science and Applied Math, Weizmann Institute of Science. u@uri.co.il. Work done while the third author was a postdoctoral fellow at the Center for Research on Computation and Society, Harvard University.

Before describing our new results, we define our setting more precisely. Consider a set of $n$ users, where each user $i$ holds as input an item $x_i \in X$. We denote $S = (x_1, \ldots, x_n)$, and refer to $S$ as a "distributed database" (as it is not stored in one location, and every $x_i$ is only held locally by user $i$). We say that a domain element $x \in X$ is $\Delta$-*heavy* if its multiplicity in $S$ is at least $\Delta$, i.e., if there are at least $\Delta$ users whose input is $x$. Our goal is to identify all $\Delta$-heavy elements for $\Delta$ as small as possible. Specifically, we receive a sequence of messages—one message $y_i$ from every user $i$—at the end of which our goal is to report a list $L \subset X$ such that

1. Every $\Delta$-heavy element $x$ is in $L$, and

2. $|L| = O(n/\Delta)$, i.e., the list size is proportional to the maximum possible number of $\Delta$-heavy elements.

We also want, for each $x \in L$, to output a value which is within $\Delta$ of the multiplicity of $x$ in $S$. The parameter $\Delta$ is referred to as *the error* of the protocol, as elements with multiplicities less than $\Delta$ are "missed" by the protocol. We refer to the probability that the protocol does not successfully achieve the above objectives as its *failure probability*, denoted by $\beta$. The privacy requirement is that for every user $i$, the distribution of the message $y_i$ should not be significantly affected by the input of the user. Formally,

**Definition 1.1** ([9])**.** *An algorithm $\mathcal{A} : X \to Y$ satisfies $\varepsilon$-differential privacy if for every two inputs $x, x'$ and for every output $y \in Y$ we have $\Pr[\mathcal{A}(x) = y] \leq e^\varepsilon \cdot \Pr[\mathcal{A}(x') = y]$.*

Solving the heavy-hitters under LDP becomes harder for smaller choices of $\Delta$,[1] and the work on LDP heavy-hitters have focused on constructing protocols that can achieve as small a value of $\Delta$ as possible.

In this work we combine ideas from the recent locally-private heavy-hitters algorithm of [3] with the recent non-private algorithm of [22], and present a new efficient LDP heavy-hitters algorithm achieving optimal worst-case error $\Delta$ as a function of the number of users $n$, the size of the domain $|X|$, the privacy parameter $\varepsilon$, and the failure probability $\beta$. Prior state-of-the-art results [4, 3] either obtained error with sub-optimal dependence on the failure probability $\beta$, or had runtime at least linear in $|X|$ (which is not realistic in real applications, where $X$ may be the space of all reasonable-length URL domains). Our result improves the worst-case error, while keeping all other complexities the same. See Table 1.

---

[1]For intuition, observe that we cannot hope to have $\Delta = 1$. Indeed, if only one user is holding some $x^*$ as input (so $x^*$ is 1-heavy), then the outcome distribution of the protocol remains approximately the same even when this user holds a different input (in which case $x^*$ does not appear in the data at all, and it is no longer 1-heavy).

| **Performance metric** | This work | Bassily et al. [3] | Bassily and Smith [4][2] |
|---|---|---|---|
| Server time | $\tilde{O}(n)$ | $\tilde{O}(n)$ | $\tilde{O}(n^{2.5})$ |
| User time | $\tilde{O}(1)$ | $\tilde{O}(1)$ | $\tilde{O}(n^{1.5})$ |
| Server memory | $\tilde{O}(\sqrt{n})$ | $\tilde{O}(\sqrt{n})$ | $\tilde{O}(n^2)$ |
| User memory | $\tilde{O}(1)$ | $\tilde{O}(1)$ | $\tilde{O}(n^{1.5})$ |
| Communication/user | $O(1)$ | $O(1)$ | $O(1)$ |
| Public randomness/user | $\tilde{O}(1)$ | $\tilde{O}(1)$ | $\tilde{O}(n^{1.5})$ |
| Worst-case error | $O\left(\frac{1}{\varepsilon} \cdot \sqrt{n \log\left(\frac{|X|}{\beta}\right)}\right)$ | $O\left(\frac{1}{\varepsilon} \cdot \sqrt{n \log\left(\frac{|X|}{\beta}\right)} \log\left(\frac{1}{\beta}\right)\right)$ | $O\left(\frac{\log^{1.5}\left(\frac{1}{\beta}\right)}{\varepsilon} \cdot \sqrt{n \log |X|}\right)$ |

Table 1: Comparison of our protocol with prior state-of-the-art results. For simplicity, the $\tilde{O}$ notation hides logarithmic factors in $n$ and $|X|$.

## 1.1 Lower Bound on the Error and New Understandings of the Local Model

We strengthen a lower bound on the error by [6, 17, 34] and [4] to incorporate the failure probability, and show that our new upper bound is tight. Our lower bound is based (conceptually) on the following new understandings of the local privacy model:

**Advanced Grouposition.** An important property of differential privacy is *group privacy*. Recall that differential privacy ensures that no single individual has a significant effect on the outcome (distribution) of the computation. A similar guarantee also holds for every small *group* of individuals, but the privacy guarantee degrades (gracefully) with the size of the group. Specifically, if an algorithm satisfies an individual-level guarantee of $\varepsilon$-differential privacy, then it satisfies $k\varepsilon$-differential privacy for every group of $k$ individuals. We observe that in the local model, group privacy degrades the privacy parameter only by a factor of $\approx \sqrt{k}$, unlike in the centralized model. This observation allows us to show improved bounds on the *max-information*—a measure of the dependence of an algorithm's output on a randomly chosen input—of locally-private protocols [8]. Strong group privacy is, however, a mixed blessing as it also leads to stronger lower bounds (of the type known as "packing arguments") for pure-private local algorithms.

**Pure vs. Approximate LDP.** Another important property of differential privacy is that it also degrades gracefully under *composition*. This allows us to design algorithms that access their input database using several (say $k$) differentially private mechanisms as subroutines, and argue about the overall privacy guarantees. In the case of pure-privacy, the privacy parameter deteriorates by a factor of at most $k$. However, in the case of approximate-privacy, $k$-fold composition degrades the privacy parameter only by a factor of $\approx \sqrt{k}$ (see [11]).

We show that in the non-interactive local model we can enjoy the best of both worlds. First, we can show that in some cases, (an approximate version of) the composition of locally private algorithms can satisfy *pure*-privacy while enjoying the same strong composition guarantees of approximate-privacy. Motivated by this observation, we then proceed to show that in the local model approximate-privacy is actually never more useful than pure-privacy (at least for non-interactive protocols). Specifically, building on a transformation of Bassily and Smith [4], we give a generic transformation from any non-interactive approximate-private local protocol into a pure-private local protocol with the same utility guarantees. This is the first formal proof that approximate local-privacy cannot enable more accurate analyses than pure local-privacy.

---

[2]The *user's* run-time and memory in [4] can be improved to $O(n)$ if one assumes random access to the public randomness, which we do not assume in this work. In fact, our protocol can be implemented without public randomness while attaining essentially the same performances.

In Section 7 we describe how our new lower bound for the heavy hitters under *pure*-LDP follows from advanced grouposition, and how the lower bound extends to *approximate*-LDP via our generic transformation. In addition, we provide a direct analysis for our lower bound that results from a direct extension of the argument of [6, 17, 34] (using ideas from [4]).

## 2 Preliminaries from Differential Privay

**Notations.** Databases are (ordered) collections of elements from some data universe $X$. Two databases $S, S' \in X^n$ are called *neighboring* if they differ in at most one entry. Throughout this paper, we use the notation $\tilde{O}$ to hide logarithmic factors in $n$ and $|X|$ (the number of users and the size of the domain).

Consider a database where each entry contains information pertaining to one individual. An algorithm operating on databases is said to preserve differential privacy if a change of a single record of the database does not significantly change the output distribution of the algorithm. Intuitively, this means that individual information is protected: whatever is learned about an individual could also be learned with her data arbitrarily modified (or without her data at all).

**Definition 2.1** (Differential Privacy [9])**.** *A randomized algorithm $M : X^n \to Y$ is $(\varepsilon, \delta)$ differentially private if for every two neighboring datasets $S, S' \in X^n$ and every $T \subseteq Y$ we have $\Pr[M(S) \in T] \leq e^\varepsilon \cdot \Pr[M(S') \in T] + \delta$, where the probability is over the randomness of $M$.*

### 2.1 Local Differential Privacy

In the local model, each individual holds her private information locally and only releases the outcomes of privacy-preserving computations on her own data. This is modeled by letting the algorithm access each entry $x_i$ in the input database $S = (x_1, \ldots, x_n) \in X^n$ separately, via differentially private *local randomizers*.

**Definition 2.2** (Local Randomizer, LR Oracle [9, 19])**.** *A local randomizer $R : X \to W$ is a randomized algorithm that takes a database of size $n = 1$. Let $S = (x_1, \ldots, x_n) \in X^n$ be a database. An LR oracle $LR_S(\cdot, \cdot)$ gets an index $i \in [n]$ and a local randomizer $R$, and outputs a random value $w \in W$ chosen according to the distribution $R(x_i)$.*

**Definition 2.3** (Local differential privacy [9, 19])**.** *An algorithm satisfies $(\varepsilon, \delta)$-local differential privacy (LDP) if it accesses the database $S$ only via the oracle $LR_S$ with the following restriction: for all possible executions of the algorithm and for all $i \in [n]$, if $LR_S(i, R_1), \ldots, LR_S(i, R_k)$ are the algorithm's invocations of $LR_S$ on index $i$, then the algorithm $\mathcal{B}(x) = (R_1(x), R_2(x), \ldots, R_k(x))$ is $(\varepsilon, \delta)$-differentially private.*

In the above definition, $k$ is the number of invocations of $LR_S$ on index $i$ made throughout the execution (potentially this number can be different for different users, and can even be chosen adaptively throughout the execution). In our heavy-hitters protocol we will have $k = 1$, meaning that every user only sends one response to the server. Local algorithms that prepare all their queries to $LR_S$ before receiving any answers are called *non-interactive*; otherwise, they are *interactive*. The focus of this work is on non-interactive protocols.

## 3 A Heavy-Hitters Protocol with Optimal Error

Let $S = (x_1, \ldots, x_n) \in X^n$ be a database, which may be distributed across $n$ users (each holding one row). For $x \in X$, we will be interested in estimating the the frequency (or multiplicity) of $x$ in $S$, i.e.,
$$f_S(x) = |\{i \in [n] : x_i = x\}|.$$
Specifically, we would like to solve the following problem while guaranteeing LDP:

**Definition 3.1.** *Let $S \in X^n$ be a (distributed) database. In the* heavy-hitters *problem with error $\Delta$ and failure probability $\beta$, the goal is to output a list* $\mathrm{Est} \subseteq X \times \mathbb{R}$ *of elements and estimates, of size $|\mathrm{Est}| = O(n/\Delta)$, such that with probability $1 - \beta$*

1. *For every $(x, a) \in \mathrm{Est}$ we have that $|f_S(x) - a| \le \Delta$; and,*

2. *For every $x \in X$ s.t. $f_S(x) \ge \Delta$ we have that $x$ appears in* $\mathrm{Est}$.

As in previous works, our algorithm for the heavy-hitters problem is based on a reduction to the simpler task of constructing a *frequency oracle*:

**Definition 3.2.** *Let $S \in X^n$ be a (distributed) database. A* frequency oracle *with error $\Delta$ and failure probability $\beta$ is an algorithm that, with probability at least $1 - \beta$, outputs a* data structure *capable of approximating $f_S(x)$ for every $x \in X$ with error at most $\Delta$.*

Observe that constructing a frequency oracle is an easier task than solving the heavy-hitters problem, as every heavy-hitters algorithm is in particular a frequency oracle. Specifically, given a solution Est to the heavy-hitters problem, we can estimate the frequency of every $x \in X$ as $\hat{f}_S(x) = a$ if $(x, a) \in \mathrm{Est}$, or $\hat{f}_S(x) = 0$ otherwise. Moreover, observe that ignoring runtime, the two problems are identical (since we can query the frequency oracle on every domain element to find the heavy-hitters).

The literature on heavy-hitters and frequency oracles under LDP has focused on the goal of minimizing the error $\Delta$. However, as Bassily and Smith showed, under LDP, every algorithm for either task must have worst case error at least
$$\Omega\left(\frac{1}{\varepsilon}\sqrt{n \cdot \log |X|}\right).$$

This lower bound is known to be tight, as constructions of frequency oracles with matching error were presented in several works [24, 17, 4, 3]. These works also presented reductions that transform a frequency oracle into a heavy-hitters algorithm. However, their reductions resulted in a sub-optimal dependency of the error in the failure probability $\beta$ (on which the lower bound above is not informative). In this work we give a new (privacy preserving) reduction from the heavy-hitters problem to the frequency oracle problem, achieving error which is optimal in all parameters. Our reduction is a private variant of the recent non-private algorithm of [22], and our final construction uses the frequency oracle of [3].

## 3.1 Existing Techniques

In this section we give an informal survey of the techniques of [3] and [22], and highlight some of the ideas behind the constructions. This intuitive overview is generally oversimplified, and hides many of the difficulties that arise in the actual analyses.

### 3.1.1 Reduction with Sub-Optimal Dependence on the Failure Probability

Assume that we have a frequency oracle protocol with worst-case error $\Delta$, and let $S = (x_1, \ldots, x_n) \in X^n$ be a (distributed) database, where user $i$ holds the input $x_i$. We now want to use our frequency oracle to identify a small set of "potential heavy-hitters". Specifically, we would like to identify all elements whose frequencies in $S$ are at least $2\Delta$. We outline the approach of [3].

Let $h : X \to [Y]$ be a (publicly known) random hash function, mapping domain elements into some range of size $Y$. We will now use $h$ in order to identify the heavy-hitters. To that end, let $x^* \in X$ denote such a heavy-hitter, appearing at least $2\Delta$ times in the database $S$, and denote $y^* = h(x^*)$. Assuming that $Y$ is big enough, w.h.p. we will have that $x^*$ is the only input element (from $S$) that is mapped (by $h$) to the hash value $y^*$. Assuming that this is indeed the case, we will now attempt to identify $x^*$ one "coordinate" (as defined below) at a time.

Let us assume that elements from $X$ are represented as $M$ symbols from an alphabet $[W]$. For every coordinate $m \in [M]$, denote $S_m = (h(x_i), x_i[m])_{i \in [n]}$, where $x_i[m]$ is the $m^{\text{th}}$ symbol of $x_i$. That is, $S_m$ is a database over the domain $([Y] \times [W])$, where the row corresponding to user $i$ is $(h(x_i), x_i[m])$. Observe that since $h$ is public, every user can compute her own row locally. As $x^*$ is a heavy-hitter, for every $m \in [M]$ we have that $(y^*, x^*[m])$ appears in $S_m$ at least $2\Delta$ times. On the other hand, as we assumed that $x^*$ is the only input element that is mapped to $y^*$, we get that $(y^*, w')$ does not appear in $S_m$ at all for every $w' \neq x^*[m]$. Recall that our frequency oracle has error at most $\Delta$, and hence, we can use it to accurately determine the symbols of $x^*$. Using this strategy we can identify a set of size $Y$ containing *all* heavy-hitters: For every hash value $y \in [Y]$, construct a potential heavy-hitter $\hat{x}^{(y)}$ symbol-by-symbol, where the $m^{\text{th}}$ symbol of $\hat{x}^{(y)}$ is the symbol $w$ s.t. $(y, w)$ is most frequent in $S_m$. By the above discussion, this strategy identifies all of the heavy-hitters.

Recall that we assumed here that $Y$ (the size of the range of the hash function $h$) is big enough so that there are no collisions among input elements (from $S$). In fact, the analysis can withstand a small number of collisions (roughly $\sqrt{n}$, since we are aiming for error bigger than $\sqrt{n}$ anyway). Nevertheless, for the above strategy to succeed with high probability, say w.p. $1 - \beta$, we will need to set $Y \gtrsim \sqrt{n}/\beta$. As $\beta$ can be exponentially small, this is unacceptable, and hence, Bassily et al. [3] applied the above strategy with a large failure probability $\beta$, and amplified the success probability using repetitions. Unfortunately, using repetitions still causes the error to increase by a factor of $\sqrt{\log(1/\beta)}$ (because applying multiple private computations to the same dataset degrades privacy). Specifically, Bassily et al. [3] obtained the following result.

**Theorem 3.3** ([3]). *Let $\varepsilon \leq 1$, and let $S \in X^n$ be a (distributed) database. There exists an $\varepsilon$-LDP algorithm that returns a list* Est *of length $\tilde{O}(\sqrt{n})$ such that with probability $1 - \beta$*

1. For every $(x, a) \in \text{Est}$ we have that $|a - f_S(x)| \leq O\left(\frac{1}{\varepsilon}\sqrt{n \log\left(\frac{\min\{n, |X|\}}{\beta}\right)}\right)$.

2. For every $x \in X$ s.t. $f_S(x) \geq O\left(\frac{1}{\varepsilon}\sqrt{n \log\left(\frac{|X|}{\beta}\right) \log\left(\frac{1}{\beta}\right)}\right)$, we have that $x$ appears in $\text{Est}$.
The server uses processing memory of size $\tilde{O}(\sqrt{n})$, and runs in time $\tilde{O}(n)$. Each user has $\tilde{O}(1)$ runtime, memory, and communication.

### 3.1.2 A Non-Private Reduction Based on List-Recoverable Codes [22]

Recall that in the reduction of Bassily et al. [3] there is only one hash function $h$. If that hash function "failed" for some heavy-hitter $x$ (meaning that too many other input elements collide with $x$), then we cannot identify $x$ and we needed to repeat. Suppose instead that we modify the reduction to introduce an *independent* hash function $h_m$ for every coordinate $m \in [M]$. The upside is that now, except with exponentially small probability (in $M$), every heavy-hitter $x^*$ causes at most a small fraction of the hash functions to fail. Then hopefully we are able to detect *most* of the symbols of $x^*$. If instead of applying this scheme to $x^*$ itself, we apply it to an error-correcting encoding of $x^*$ with constant rate and constant relative distance, then we can indeed recover $x^*$ this way.[3] However, aside from the fact that some of the symbols are missing, it now becomes unclear how to concatenate the coordinates of the elements that we are constructing. Specifically, in the construction of Bassily et al., the fact that $h(x^*) = y^*$ appeared throughout all coordinates is what we used to concatenate the different symbols we identified. If the hash functions are independent across $m$, then this does not work.

We now describe this strategy in more detail, and describe the solution proposed by Larsen et al. [22] to overcome this issue. Fix a heavy-hitter $x^*$. For every coordinate $m \in [M]$ where $h_m$ does not fail on $x^*$, we have that the frequency of $(h_m(x^*), x^*[m])$ "stands out" in comparison to every $(h_m(x^*), w')$ where $w' \neq x^*[m]$, meaning that $(h_m(x^*), x^*[m])$ appears in $S_m$ significantly more times than any other $(h_m(x^*), w')$. Hence, for every coordinate $m \in [M]$ and for every hash value $y \in [Y]$ we identify (at most 1) symbol $w$ s.t. $(y, w)$ "stands out" compared to all other $(y, w')$ in $S_m$. In other words, for every $m \in [M]$ we construct a list $L_m = \{(y, w)\}$ of "stand outs" in $S_m$. The guarantee is that for every heavy-hitter $x^*$, we have that $(h_m(x^*), x^*[m])$ appears in $L_m$ for most coordinates $m \in [M]$. If we could single out all occurrences of $(h_m(x^*), x^*[m])$ in these lists, then we could concatenate the symbols $x^*[m]$ and reconstruct $x^*$. The question is how to match symbols from different lists. As was observed in [14] that studied the related problem of $\ell_1/\ell_1$ compressed sensing, this can be done using *list-recoverable codes*.

Recall that in a (standard) error correcting code, the decoder receives a noisy codeword (represented by, say, $M$ symbols from $[W]$) and recovers a (noiseless) codeword that agrees with the noisy codeword on at least $(1 - \alpha)$ fraction of the coordinates, assuming that such a codeword exists. An error correcting code thus protects a transmission from any adversarial corruption of an $\alpha$-fraction of a codeword. A *list-recoverable code* protects a transmission from uncertainty as well as corruption. Instead of receiving a single (possibly corrupted) symbol in each coordinate, the decoder for a list-recoverable code receives a short list of size $\ell$ of possible symbols. Let $L_1, \ldots, L_M$ be such a sequence of lists, where list $L_m$ contains the $m^{\text{th}}$ symbol of a noisy codeword. The requirement on

---
[3]We remark that this presentation is oversimplified. In the analysis of [22], the problem is not only that each $x^*$ causes a small fraction of the hash functions to fail. Rather, the problem is that a small fraction of the coordinates are "bad" (since otherwise it may be that each $x$ is failed by very few hash functions, but every $h_m$ ends up "bad" because they fail for some different $x$'s).

the decoder is that it is able to recover all possible codewords $x$ that "agree" with at least $(1 - \alpha)$ fraction of the lists, in the sense that $m^{\text{th}}$ symbol of $x$ appears in $L_m$. Formally,

**Definition 3.4** ([16]). *An $(\alpha, \ell, L)$-list-recoverable code is a pair of mappings* (Enc, Dec) *where* Enc : $X \to [Z]^M$, *and* Dec : $([Z]^\ell)^M \to X^L$, *such that the following holds. Let* $L_1, \ldots, L_M \in [Z]^\ell$. *Then for every* $x \in X$ *satisfying* $|\{m : \text{Enc}(x)_m \in L_m\}| \geq (1-\alpha)M$ *we have that* $x \in \text{Dec}(L_1, \ldots, L_M)$.

In our case, the lists $L_m = \{(y, w)\}$ contain the "stand outs" of $S_m$, with the guarantee that for every heavy-hitter $x^*$, we have that $(h_m(x^*), x^*[m])$ appears in $L_m$ for most coordinates $m$. We can therefore, in principle, use a list-recoverable code in order to recover all heavy-hitters (in fact, with this formulation it suffices that the lists $L_m$ contain only the set of heavy items $\{w\}$ found to be heavy in $S_m$). However, there are currently no explicit codes achieving the parameter settings that are needed for the construction. To overcome this issue, Larsen et al. [22] constructed a *relaxed* variant of list-recoverable codes, and showed that their relaxed variant still suffices for their (non-private) reduction. Intuitively, their relaxed code utilizes the randomness in the first portion of every element in the lists (i.e., the outcome of the hash functions). Formally,

**Definition 3.5.** *An $(\alpha, \ell, L)$-unique-list-recoverable code is a pair of mappings* (Enc, Dec) *where* Enc : $X \to ([Y] \times [Z])^M$, *and* Dec : $(([Y] \times [Z])^\ell)^M \to X^L$, *such that the following holds. Let* $L_1, \ldots, L_M \in ([Y] \times [Z])^\ell$. *Assume that for every* $m \in [M]$, *if* $(y, z), (y', z') \in L_m$ *then* $y \neq y'$. *Then for every* $x \in X$ *satisfying* $|\{m : \text{Enc}(x)_m \in L_m\}| \geq (1-\alpha)M$ *we have that* $x \in \text{Dec}(L_1, \ldots, L_M)$.

**Theorem 3.6** ([22]). *There exist constants $C > 1$ and $0 < \alpha < 1$ such that the following holds. For all constants $M \leq \log |X|$ and $Y, \ell \in \mathbb{N}$, and for every fixed choice of functions $h_1, \ldots, h_M : X \to [Y]$, there is a construction of an $(\alpha, \ell, L)$-unique-list-recoverable code*

$$\text{Enc} : X \to ([Y] \times [Z])^M \qquad \text{and} \qquad \text{Dec} : (([Y] \times [Z])^\ell)^M \to X^L,$$

*where $L \leq C \cdot \ell$ and $Z \leq (|X|^{1/M} \cdot Y)^C$. Furthermore, there is a mapping $\widetilde{\text{Enc}} : X \to [Z]^M$ such that for every $x \in X$ we have $\text{Enc}(x) = \left( (h_1(x), \widetilde{\text{Enc}}(x)_1), \ldots, (h_M(x), \widetilde{\text{Enc}}(x)_M) \right)$.*

*There is a pre-processing algorithm which, given $M, Y, \ell, h_1, \ldots, h_M$, computes a description of the functions* Enc, Dec *in* poly($M$) *time.[4] Then, evaluating* Enc($x$) *takes linear time and space in addition to $O(M)$ invocations of the hash functions, and evaluating* Dec($L_1, \ldots, L_M$) *takes linear space and polynomial time.*

For completeness, we include the proof in Appendix B. In the following sections we construct a private version of the algorithm of [22] and show that it achieves optimal error in all parameters. We remark that the focus in [22] was on space and runtime, and that in fact, in the non-private literature the meaning of "minimal achievable error" is unclear, as achieving zero error is trivial (it is "only" a question of runtime and memory).

## 3.2 Additional Preliminaries

We start by presenting the additional preliminaries that enable our construction.

---
[4]The pre-processing algorithm has deterministic poly($M$) time for every $M$ of the form $M = D^i$ for some universal constant $D$. Otherwise, the algorithm has poly($M$) Las Vegas expected runtime. We ignore this issue for simplicity.

### 3.2.1 Frequency Oracle

Recall that a frequency oracle is a protocol that, after communicating with the users, outputs a *data structure* capable of approximating the frequency of every domain element $x \in X$. Our protocol uses the frequency oracle of [3], named `Hashtogram`, as a subroutine.

**Theorem 3.7** (Algorithm `Hashtogram` [3]). *Let $S \in X^n$ be a database which is distributed across $n$ users, and let $\beta, \varepsilon \leq 1$. There exists an $\varepsilon$-LDP algorithm such that the following holds. Fix a domain element $x \in X$ to be given as a query to the algorithm. With probability at least $1 - \beta$, the algorithm answers the query $x$ with $a(x)$ satisfying:*

$$|a(x) - f_S(x)| \leq O\left(\frac{1}{\varepsilon} \cdot \sqrt{n \log\left(\frac{\min\{n, |X|\}}{\beta}\right)}\right).$$

*The server uses processing memory of size $\tilde{O}(\sqrt{n})$, and runs in time $\tilde{O}(1)$ per query (plus $\tilde{O}(n)$ for instantiation). Each user has $\tilde{O}(1)$ runtime, memory, and communication. The $\tilde{O}$ notation hides logarithmic factors in $n$ and $|X|$.*

Observe that by a union bound, algorithm `Hashtogram` answers every fixed set of $w$ queries with worst case error $O\left(\frac{1}{\varepsilon} \cdot \sqrt{n \log\left(\frac{w \cdot \min\{n, |X|\}}{\beta}\right)}\right)$. The $\min\{n, |X|\}$ factor is known to be unnecessary for some regimes of $n, |X|, \beta$, or if one allows a slightly larger runtime or memory. In particular, Bassily et al. gave a (simplified) analysis of algorithm `Hashtogram` for the case where the domain $X$ is small, with the following guarantees.

**Theorem 3.8** (Algorithm `Hashtogram` [3]). *Let $S \in X^n$ be a database which is distributed across $n$ users, and let $\beta, \varepsilon \leq 1$. There exists an $\varepsilon$-LDP algorithm such that the following holds. Fix a domain element $x \in X$ to be given as a query to the algorithm. With probability at least $1 - \beta$, the algorithm answers the query $x$ with $a(x)$ satisfying:*

$$|a(x) - f_S(x)| \leq O\left(\frac{1}{\varepsilon} \cdot \sqrt{n \log\left(\frac{1}{\beta}\right)}\right).$$

*The server uses processing memory of size $\tilde{O}(|X|)$, and runs in time $\tilde{O}(1)$ per query (plus $\tilde{O}\left(|X|^2\right)$ for instantiation). Each user has $\tilde{O}(1)$ runtime, memory, and communication. The $\tilde{O}$ notation hides logarithmic factors in $n$ and $|X|$.*

We will only apply Theorem 3.8 on small domains $X$ (satisfying $|X|^2 = \tilde{O}(n)$). Our eventual analysis in the proof of Theorem 3.13 will make use of both accuracy guarantees of `Hashtogram`.

### 3.2.2 The Poisson Approximation

When throwing $n$ balls into $R$ bins, the distribution of the number of balls in a given bin is $\text{Bin}(n, 1/R)$. As the Poisson distribution is the limit distribution of the binomial distribution when $n/R$ is fixed and $n \to \infty$, the distribution of the number of balls in a given bin is approximately $\text{Pois}(n/R)$. In fact, in some cases we may approximate the *joint distribution* of the number of balls in all the bins by assuming that the load in each bin is an independent Poisson random variable with mean $n/R$.

**Theorem 3.9** (e.g., [25])**.** *Suppose that $n$ balls are thrown into $R$ bins independently and uniformly at random, and let $X_i$ be the number of balls in the $i^{th}$ bin, where $1 \leq i \leq R$. Let $Y_1, \cdots, Y_R$ be independent Poisson random variables with mean $n/R$. Let $f(x_1, \cdots, x_R)$ be a nonnegative function. Then,*

$$\mathbb{E}\left[f(X_1, \cdots, X_R)\right] \leq e\sqrt{n}\,\mathbb{E}\left[f(Y_1, \cdots, Y_R)\right].$$

In particular, the theorem states that any event that takes place with probability $p$ in the Poisson case, takes place with probability at most $pe\sqrt{n}$ in the exact case (this follows by letting $f$ be the indicator function of that event).

We will also use the following bounds for the tail probabilities of a Poisson random variable:

**Theorem 3.10** ([2])**.** *Let $X$ have Poisson distribution with mean $\mu$. For $0 \leq \alpha \leq 1$,*

$$\begin{aligned}
\Pr[X \leq \mu(1-\alpha)] &\leq e^{-\alpha^2 \mu/2} \\
\Pr[X \geq \mu(1+\alpha)] &\leq e^{-\alpha^2 \mu/2}.
\end{aligned}$$

### 3.2.3 Concentration (with Limited Independence)

We first state a version of the multiplicative Chernoff bound, including a formulation of the upper tail bound which holds under limited independence.

**Theorem 3.11** ([30, Theorem 2])**.** *Let $X_1, \ldots, X_n$ be random variables taking values in $\{0, 1\}$. Let $X = \sum_{i=1}^n X_i$ and $\mu = \mathbb{E}[X]$. Then for every $0 \leq \alpha \leq 1$,*

1. *If the $X_i$'s are $\lceil \mu\alpha \rceil$-wise independent, then $\Pr[X \geq \mu(1+\alpha)] \leq \exp(-\alpha^2 \mu/3)$.*

2. *If the $X_i$'s are fully independent, then $\Pr[X \leq \mu(1-\alpha)] \leq \exp(-\alpha^2 \mu/2)$.*

Next, we present a limited independence version of Bernstein's inequality.

**Theorem 3.12** ([18, Lemma 2])**.** *Let $X_1, \ldots, X_n$ be $k$-wise independent random variables, for an even integer $k \geq 2$, that are each bounded by $T$ in magnitude. Let $X = \sum_{i=1}^n X_i$ and $\mu = \mathbb{E}[X]$. Let $\sigma^2 = \sum_i \mathbb{E}(X_i - \mathbb{E}\,X_i)^2$. Then there exists a constant $C > 0$ such that for all $\lambda > 0$,*

$$\Pr\left[|X - \mu| > \lambda\right] \leq C^k \cdot \left((\sigma\sqrt{k}/\lambda)^k + (Tk/\lambda)^k\right).$$

## 3.3 The Full Protocol

In this section we construct and analyze our heavy-hitters algorithm. The full algorithm appears as Algorithm `PrivateExpanderSketch`.

It is immediate from the construction that algorithm `PrivateExpanderSketch` satisfies $\varepsilon$-LDP. We now proceed with the utility analysis. While we must make the restriction $n \leq |X|$ for the analysis to go through, we remark that in the complementary case $n > |X|$, it is possible to just apply the frequency oracle `Hashtogram` (Theorem 3.7) to every item in $X$ and obtain the same runtime, memory, and communication guarantees.

**Algorithm** `PrivateExpanderSketch`

**Notation:** Let $C_M, C_Y, C_\ell, C_g,$ and $C_f$ be universal constants (to be specified later).
Denote $M = C_M \cdot \log|X|/\log\log|X|$, $Y = \log^{C_Y}|X|$, and $\ell = C_\ell \cdot \log|X|$.

**Public randomness:** Random partition of $[n]$ into $M$ sets $I_1, \ldots, I_M$.
Pairwise independent hash functions $h_1, \cdots, h_M : X \to [Y]$.
$(C_g \cdot \log|X|)$-wise independent hash function $g : X \to [B]$.

**Tool used:** A unique-list-recoverable code $(\text{Enc}, \text{Dec})$ with parameters $M, Y, \ell$ using $h_1, \ldots, h_M$, as in Theorem 3.6.

**Setting:** Each player $i \in [n]$ holds a value $x_i \in X$. Define $S = (x_1, \cdots, x_n)$.
For $m \in [M]$, let $S_m = \left(g(x_i), \text{Enc}(x_i)_m\right)_{i \in I_m} = \left(g(x_i), h_m(x_i), \widetilde{\text{Enc}}(x_i)_m\right)_{i \in I_m}$.

1. For every $m \in [M]$, use $\texttt{Hashtogram}(S_m)$ with privacy parameter $\frac{\varepsilon}{2}$ to get estimates $\hat{f}_{S_m}(b,y,z)$ for $f_{S_m}(b,y,z)$ for every $(b,y,z) \in [B]\times[Y]\times[Z]$.

2. For every $(m,b) \in [M]\times[B]$, initialize $L_m^b = \emptyset$.

3. For every $(m,b,y) \in [M]\times[B]\times[Y]$:

    (a) Let $z = \operatorname{argmax}\{\hat{f}_{S_m}(b,y,z)\}$.

    (b) If $\hat{f}_{S_m}(b,y,z) \geq C_f \cdot \frac{\log\log|X|}{\varepsilon} \cdot \sqrt{\frac{n}{\log|X|}} = O\left(\frac{\sqrt{n\log|X|}}{\varepsilon M}\right)$ and $|L_m^b| \leq C_\ell \cdot \log|X|$ then add $(y,z)$ to $L_m^b$.

4. For every $b \in [B]$ let $\widehat{H}^b = \text{Dec}(L_1^b, \ldots, L_M^b)$. Define $\widehat{H} = \bigcup_{b \in [B]} \widehat{H}^b$.

5. Use $\texttt{Hashtogram}(S)$ with privacy parameter $\frac{\varepsilon}{2}$ to obtain $\hat{f}_S(x) \approx f_S(x)$ for every $x \in \widehat{H}$.

6. Return $\text{Est} = \left\{\left(x, \hat{f}_S(x)\right) : x \in \widehat{H}\right\}$.

**Theorem 3.13.** *Let $S \in X^n$ be a database which is distributed across $n$ users, with $n \leq |X|$, and let $\beta, \varepsilon \leq 1$. Algorithm `PrivateExpanderSketch` returns a list $\text{Est}$ of length $\tilde{O}(\sqrt{n})$ s.t. with probability $1 - \beta$ we have that*

1. *For every $(x,a) \in \text{Est}$ we have that $|a - f_S(x)| \leq O\left(\frac{1}{\varepsilon}\sqrt{n\log\left(\frac{\min\{n,|X|\}}{\beta}\right)}\right)$.*

2. *For every $x \in X$ s.t. $f_S(x) \geq O\left(\frac{1}{\varepsilon}\sqrt{n\log\left(\frac{|X|}{\beta}\right)}\right)$, we have that $x$ is included in $\text{Est}$.*

*The server uses processing memory of size $\tilde{O}(\sqrt{n})$, and runs in time $\tilde{O}(n)$. Each user has $\tilde{O}(1)$ runtime, memory, and communication.*

*Proof.* Item 1 of the lemma follows directly from Theorem 3.7 (the properties of `Hashtogram`). We now prove item 2. Below, we will freely assume that $n \geq O(\log|X|/\varepsilon^2)$, with item 2 being vacuously true otherwise. For some constant $C_H \geq 1$, we say that an element $x \in X$ is *heavy* if $f_S(x) \geq \frac{C_H}{\varepsilon}\sqrt{n \cdot \log|X|}$, and *non-heavy* otherwise. We let $H$ denote the set of all heavy elements. Observe that $|H| \leq \varepsilon \cdot \sqrt{n/\log|X|}$. For $b \in [B]$ denote the set of all heavy elements that are mapped to $b$ under $g$ as $H_b = \{x \in H : g(x) = b\}$.

We begin by defining a sequence of events $E_1, \ldots, E_7$. We will show that all of these events occur simultaneously with high probability, and that if they indeed all occur, then the guarantee of item 2 holds. First consider the following event:

> **Event $E_1$ (over the choice of $g$):**
> For every $b \in [B]$, the number of heavy elements that are mapped to $b$ under $g$ is at most $C_1 \log |X|$. Furthermore, the sum of frequencies of all non-heavy elements (i.e., elements $x \notin H$) that are mapped to $b$ is at most $\frac{C_1}{\varepsilon} \sqrt{n} \log^{3/2} |X|$, where $C_1 > 0$ is an absolute constant. That is, $\forall b \in [B]$:
> $$|H_b| \leq C_1 \log |X| \qquad \text{and} \qquad \sum_{x \notin H:\ g(x)=b} f_S(x) \leq \frac{C_1}{\varepsilon} \sqrt{n} \log^{3/2} |X|$$

We now show that $E_1$ happens with high probability. For $B = \left\lceil \frac{1}{10 C_H} \cdot \varepsilon \sqrt{n} / \log^{3/2} |X| \right\rceil \leq |X|$, it follows by the limited independence Chernoff bound (Theorem 3.11, item 1) and a union bound over all $b \in [B]$ that for some constant $C_1 > 0$,

$$\Pr_g \left[ \exists b \in [B] \text{ s.t. } |\{x \in H : g(x) = b\}| > C_1 \log |X| \right] < \frac{1}{2|X|}. \tag{1}$$

Here, we made use of the fact that $g: X \to [B]$ is $(C_g \cdot \log |X|)$-wise independent for a constant $C_g$. We next invoke the limited independence formulation of Bernstein's inequality (Theorem 3.12). Fix $b \in [B]$. We will consider the collection of random variables $R_x$ indexed by $x \in X \setminus H$ defined as follows. Define $R_x = \mathbf{1}_{g(x)=b} \cdot f_S(x)$. Then each $R_x$ is bounded by $T < \frac{1}{\varepsilon} \sqrt{n \cdot \log |X|}$ and $\sum_{x \notin H} R_x$ has variance

$$\sigma^2 = \sum_{x \notin H} f_S^2(x) \left( \frac{1}{B} - \frac{1}{B^2} \right) \leq \sum_{x \notin H} \frac{1}{B} \cdot f_S^2(x) \leq \frac{1}{\varepsilon^2} n \log^2 |X|.$$

It then follows by Theorem 3.12 and a union bound over all $b \in [B]$ that

$$\Pr_g \left[ \exists b \in [B] \text{ s.t. } \sum_{x \notin H:\ g(x)=b} f_S(x) > \frac{C_1}{\varepsilon} \sqrt{n} \log^{3/2} |X| \right] < \frac{1}{2|X|}. \tag{2}$$

This shows that $\Pr[E_1] \geq 1 - \frac{1}{|X|}$. We now continue the analysis assuming that Event $E_1$ occurs.

> **Event $E_2$ (over partitioning $[n]$ into $I_1, \ldots, I_M$):**
> For every $(b, m) \in [B] \times [M]$ we have $|\{i \in I_m : x_i \notin H \text{ and } g(x_i) = b\}| \leq \frac{2 C_1}{M \varepsilon} \sqrt{n} \log^{3/2} |X|$.

Conditioned on Event $E_1$, for every $b \in [B]$ we have that $|\{i \in [n] : x_i \notin H \text{ and } g(x_i) = b\}| \leq \frac{C_1}{\varepsilon} \sqrt{n} \log^{3/2} |X|$. Hence, by the Chernoff bound (Theorem 3.11, item 2) and a union bound over all $(b, m) \in [B] \times [M]$, we get that $\Pr[E_2 | E_1] \geq 1 - \frac{1}{|X|}$.

> **Event $E_3$ (over partitioning $[n]$ into $I_1, \ldots, I_M$):**
> For every $x \in H$ there exists a subset $M_3^x \subseteq [M]$ of size $|M_3^x| \geq (1 - \frac{\alpha}{10}) M$ s.t. for every $m \in M_3^x$ we have that $|\{i \in I_m : x_i = x\}| \geq \frac{f_S(x)}{2M}$.

Recall that $M = C_M \cdot \log|X|/\log\log|X|$. As in Theorem 3.9 (the Poisson approximation), we analyze event $E_3$ in the Poisson case. To that end, fix $x \in H$, and let $\tilde{J}_1, \cdots, \tilde{J}_M$ be independent Poisson random variables with mean $\frac{f_S(x)}{M}$. Let us say that $m \in [M]$ is *bad* if $\tilde{J}_m < \frac{f_S(x)}{2M}$. Now fix $m \in [M]$. Using a tail bound for the Poisson distribution (see Theorem 3.10), assuming that $n \geq 8C_M^2 \log|X|$, we have that $m$ is bad with probability at most $\frac{1}{\log^{C_M}|X|}$. As $\tilde{J}_1, \cdots, \tilde{J}_M$ are independent, the probability that there are more than $\alpha M/10$ bad choices for $m$ is at most

$$\binom{M}{\alpha M/10} \left(\frac{1}{\log^{C_M}|X|}\right)^{\alpha M/10} \leq \left(\frac{10e}{\alpha \log^{C_M}|X|}\right)^{\frac{\alpha C_M \log|X|}{10 \log\log|X|}}$$

$$= \left(\frac{10e}{\alpha}\right)^{\frac{\alpha C_M \log|X|}{10 \log\log|X|}} \cdot \left(\frac{1}{|X|}\right)^{\frac{\alpha C_M^2}{10}}$$

$$= \left(\frac{1}{|X|}\right)^{\frac{\alpha C_M^2}{10} - o(1)}.$$

Hence, by Theorem 3.9 we get that

$$\Pr\left[\begin{array}{c} \text{There are at least } \alpha M/10 \text{ indices } m \text{ s.t.} \\ |\{i \in I_m : x_i = x\}| < f_S(x)/(2M) \end{array}\right] \leq e\sqrt{f_S(x)} \cdot \left(\frac{1}{|X|}\right)^{\frac{\alpha C_M^2}{10} - o(1)}.$$

Choosing $C_M \geq \frac{10}{\alpha}$ and using the union bound over every $x \in H$ we get that $\Pr[E_3] \geq 1 - \frac{1}{|X|}$.

> **Event $E_4$ (over partitioning $[n]$ into $I_1, \ldots, I_M$):**
> There exists a subset $M_4 \subseteq [M]$ of size $|M_4| \geq (1 - \frac{\alpha}{10})M$ s.t. for every $m \in M_4$ we have $\frac{n}{2M} \leq |I_m| \leq \frac{2n}{M}$.

An analysis similar to that of Event $E_3$ shows that $\Pr[E_4] \geq 1 - \frac{1}{|X|}$.

> **Event $E_5$ (over sampling $h_1, \cdots, h_M$):**
> For every $b \in [B]$ there exists a subset $M_5^b \subseteq [M]$ of size $|M_5^b| \geq (1 - \frac{\alpha}{10})M$ s.t. for every $m \in M_5^b$ we have that $h_m$ perfectly hashes the elements of $H_b = \{x \in H : g(x) = b\}$.

We analyze event $E_5$ assuming that event $E_1$ occurs, in which case $|H_b| \leq C_1 \cdot \log|X|$ for every $b \in [B]$. Fix $b \in [B]$ and $m \in [M]$. Recall that $h_m$ is a pairwise independent hash function mapping $X$ to $[Y]$. Hence, for $x' \neq x$ we have $\Pr_{h_m}[h_m(x') = h_m(x)] = \frac{1}{Y}$. Using the union bound over every pair $(x, x')$ with $x' \neq x \in H_b$, assuming that $Y \geq C_1^2 \cdot \log^{C_M+2}|X|$, we have

$$\Pr_{h_m}[\exists (x, x') \in H_b \times H_b \text{ s.t. } x \neq x', \ h_m(x') = h_m(x)] \leq \frac{1}{Y}\binom{|H_b|}{2} \leq \frac{1}{\log^{C_M}|X|}.$$

That is, the probability that $h_m$ does not perfectly hash the elements of $H_b$ is at most $\frac{1}{\log^{C_M}|X|}$. As the hash functions $h_1, \ldots, h_M$ are independent, the probability that there are more than $\alpha M/10$ choices of $m$ s.t. $h_m$ does not perfectly hash $H_b$ is at most

$$\binom{M}{\alpha M/10} \left(\frac{1}{\log^{C_M} |X|}\right)^{\alpha M/10} \leq \left(\frac{1}{|X|}\right)^{\frac{\alpha C_M^2}{10} - o(1)}.$$

Setting $C_M \geq \frac{10}{\alpha}$ and using the union bound over every $b \in [B]$ we get that $\Pr[E_5 | E_1] \geq 1 - \frac{1}{|X|}$.

> **Event $E_6$ (over sampling $h_1, \cdots, h_M$):**
> For every $x \in H$ there exists a subset $M_6^x \subseteq [M]$ of size $|M_6^x| \geq (1 - \frac{\alpha}{10})M$ s.t. for every $m \in M_6^x$ we have that $|\{i \in I_m : x_i \notin H \text{ and } g(x_i) = g(x) \text{ and } h_m(x_i) = h_m(x)\}| \leq \frac{1}{\varepsilon M}\sqrt{n \cdot \log |X|}$.

We analyze event $E_6$ assuming that event $E_2$ occurs, in which case for every $b, m$ we have that $|\{i \in I_m : x_i \notin H, \ g(x_i) = b\}| \leq \frac{2C_1}{M\varepsilon}\sqrt{n} \log^{3/2} |X|$. Fix $x \in H$ and fix $m \in [M]$. We have that

$$\mathbb{E}_{h_m}\left[|\{i \in I_m : x_i \notin H, \ g(x_i) = g(x), \ h_m(x_i) = h_m(x)\}|\right] = \sum_{\substack{i \in I_m: \\ x_i \notin H, \\ g(x_i) = g(x)}} \mathbb{E}_{h_m}\left[\mathbb{1}_{h_m(x_i) = h_m(x)}\right]$$

$$\leq \frac{1}{Y} \cdot \frac{2C_1}{M\varepsilon}\sqrt{n} \log^{3/2} |X|.$$

Thus, by Markov's inequality, we have that

$$\Pr_{h_m}\left[|\{i \in I_m : x_i \notin H, \ g(x_i) = g(x), \ h_m(x_i) = h_m(x)\}| \geq \frac{1}{\varepsilon M}\sqrt{n \cdot \log |X|}\right] \leq \frac{2C_1 \log |X|}{Y}.$$

Let us say that $m$ is *bad* for $x$ if $|\{i \in I_m : x_i \notin H, \ g(x_i) = g(x), \ h_m(x_i) = h_m(x)\}| \geq \frac{1}{\varepsilon M}\sqrt{n \cdot \log |X|}$. So $m$ is bad with probability at most $\frac{2C_1 \log |X|}{Y}$. As the hash functions $h_1, \ldots, h_M$ are independent, assuming that $Y \geq 2C_1 \cdot \log^{C_M+1} |X|$, the probability that there are more than $\alpha M/10$ bad choices for $m$ is at most

$$\binom{M}{\alpha M/10} \left(\frac{1}{\log^{C_M} |X|}\right)^{\alpha M/10} \leq \left(\frac{1}{|X|}\right)^{\frac{\alpha C_M^2}{10} - o(1)}.$$

Setting $C_M \geq \frac{10}{\alpha}$ and using the union bound over every $x \in H$ we get that $\Pr[E_6 | E_2] \geq 1 - \frac{1}{|X|}$.

> **Event $E_7$ (over the randomness of `Hashtogram`):**
> For every $b \in [B]$ there exists a subset $M_7^b \subseteq [M]$ of size $|M_7^b| \geq (1 - \frac{\alpha}{5})M$ s.t. for every $m \in M_7^b$ and every $(y, z) \in [Y] \times [Z]$ we have that $\left|\hat{f}_{S_m}(b, y, z) - f_{S_m}(b, y, z)\right| \leq \frac{C_7 \cdot \log \log |X|}{\varepsilon} \sqrt{n/\log |X|}$, where $C_7 > 0$ is an absolute constant.

We analyze Event $E_7$ conditioned on Event $E_4$, and let $M_4 \subseteq [M]$ denote the set from Event $E_4$. Recall that $|M_4| \geq (1 - \frac{\alpha}{10})M$, and that for every $m \in M_4$ we have that $n/2M \leq |S_m| \leq 2n/M$.

Fix $b \in [B]$ and fix $m \in M_4$. Let us say that $m$ is *bad* if there exists $(y, z) \in [Y] \times [Z]$ s.t. $\left|\hat{f}_{S_m}(b, y, z) - f_{S_m}(b, y, z)\right| > \frac{6}{\varepsilon}\sqrt{|S_m| \cdot \log(4YZ \cdot \log^{C_M} |X|)} = O\left(\frac{\log \log |X|}{\varepsilon} \sqrt{n/\log |X|}\right)$. By the properties of algorithm Hashtogram (Theorem 3.8), we have that $m$ is bad with probability at most $\frac{1}{\log^{C_M} |X|}$. As the coins of Hashtogram are independent across different executions (i.e., for different values of $m$), the probability that there are more than $\alpha |M_4|/10$ bad choices for $m \in M_4$ is at most

$$\binom{|M_4|}{\alpha |M_4|/10} \left(\frac{1}{\log^{C_M} |X|}\right)^{\alpha |M_4|/10} \leq \left(\frac{1}{|X|}\right)^{\frac{\alpha C_M^2}{20} - o(1)}.$$

Setting $C_M \geq \frac{20}{\alpha}$ and using the union bound over every $b \in [B]$ we get that $\Pr[E_7 | E_4] \geq 1 - \frac{1}{|X|}$.

We are now ready to complete the proof, by showing that whenever $E_1, E_2, E_3, E_4, E_5, E_6, E_7$ occur we have that every heavy element $x \in H$ appears in the output list $\widehat{H}$. To that end, fix $x \in H$, and denote $M^x = M_3^x \cap M_4 \cap M_5^{g(x)} \cap M_6^x \cap M_7^{g(x)}$. Observe that $|M^x| \geq (1 - \alpha)M$. We now show that for every $m \in M^x$ it holds that $(h_m(x), \widetilde{\text{Enc}}(x)_m) \in L_m^{g(x)}$, in which case, by the properties of the code (Enc, Dec) we have that $x \in \widehat{H}^{g(x)} = \text{Dec}(L_1^{g(x)}, \ldots, L_M^{g(x)})$, and hence, $x \in \widehat{H}$.

So let $m \in M^x$. By Event $E_3$ we have that

$$f_{S_m}\left(g(x), h_m(x), \widetilde{\text{Enc}}(x)_m\right) \geq \frac{C_H \cdot \log \log |X|}{2C_M \cdot \varepsilon} \sqrt{n/\log |X|}.$$

Hence, by Event $E_7$ we have that

$$\hat{f}_{S_m}\left(g(x), h_m(x), \widetilde{\text{Enc}}(x)_m\right) \geq C_f \cdot \frac{\log \log |X|}{\varepsilon} \sqrt{n/\log |X|}, \tag{3}$$

where $C_f = \left(\frac{C_H}{2C_M} - C_7\right)$.

On the other hand, let $z \neq \widetilde{\text{Enc}}(x)_m$, and recall that by Event $E_5$ there does not exist a heavy element $x' \neq x$ s.t. $g(x') = g(x)$ and $h_m(x') = h_m(x)$. In addition, by Event $E_6$, there could be at most $\frac{\log \log |X|}{\varepsilon} \sqrt{n/\log |X|}$ indices $i \in I_m$ s.t. $x_i$ is non-heavy and $g(x_i) = g(x)$ and $h_m(x_i) = h_m(x)$. That is,

$$f_{S_m}(g(x), h_m(x), z) = \left|\left\{i \in I_m : \begin{array}{l} x_i \in H, \\ g(x_i) = g(x), \\ h_m(x_i) = h_m(x), \\ \widetilde{\text{Enc}}(x_i)_m = z \end{array}\right\}\right| + \left|\left\{i \in I_m : \begin{array}{l} x_i \notin H, \\ g(x_i) = g(x), \\ h_m(x_i) = h_m(x), \\ \widetilde{\text{Enc}}(x_i)_m = z \end{array}\right\}\right|$$

$$\leq 0 + \frac{\log \log |X|}{\varepsilon} \sqrt{n/\log |X|}.$$

Hence, by Event $E_7$ we have that

$$\hat{f}_{S_m}(g(x), h_m(x), z) \leq (C_7 + 1) \frac{\log \log |X|}{\varepsilon} \sqrt{n/\log |X|}. \tag{4}$$

By Inequalities (3) and (4), and by setting $C_H$ large enough, we ensure that $\widetilde{\text{Enc}}(x)_m$ is identified in step 3a as the argmax of $\hat{f}_{S_m}(g(x), h_m(x), \cdot)$. Hence, in order to show that $(h_m(x), \widetilde{\text{Enc}}(x)_m)$ is added to $L_m^{g(x)}$ in step 3b, it suffices to show that $|L_m^{g(x)}| \leq C_\ell \cdot \log |X|$ for some constant $C_\ell$. First recall that, by Event $E_1$, there are at most $C_1 \cdot \log|X|$ heavy elements $x' \in H$ s.t. $g(x') = g(x)$. Hence, there could be at most $C_1 \cdot \log|X|$ pairs $(y, z) \in L_m^{g(x)}$ s.t. $\exists x' \in H$ s.t. $(g(x'), h_m(x'), \widetilde{\text{Enc}}_m(x')) = (g(x), y, z)$. Therefore, it suffices to show that $L_m^{g(x)}$ contains at most $O(\log|X|)$ pairs $(y, z)$ that are added because they do *not* correspond to a heavy element $x' \in H$.

To that end, observe that by the condition on step 3b, and by Event $E_7$, in order for a pair $(y, z)$ to be added to $L_m^{g(x)}$ it must be that $f_{S_m}(g(x), y, z) \geq \frac{\log \log |X|}{\varepsilon}\sqrt{n/\log|X|}$. However, by Event $E_2$, there are at most $\frac{2C_1 \cdot \log \log |X|}{C_M \cdot \varepsilon}\sqrt{n \log |X|}$ elements in $S_m$ that yield the same value $g(x)$, but are not generated from a heavy element. Hence, there could be at most $\frac{2C_1}{C_M} \cdot \log|X|$ pairs $(y, z) \in L_m^{g(x)}$ such that there is no $x' \in H$ where $(g(x'), h_m(x'), \widetilde{\text{Enc}}_m(x')) = (g(x), y, z)$. Overall we have that $|L_m^{g(x)}| \leq C_\ell \cdot \log |X|$ for $C_\ell = C_1 + 2C_1/C_M$. $\square$

## 4 Advanced Grouposition and Max-Information

In this section, we show that local differential privacy admits stronger guarantees of group privacy than differential privacy in the central model. In particular, the protection offered to groups of size $k$ under LDP degrades proportionally to about $\sqrt{k}$, a quadratic improvement over what happens in the central model. It is not a coincidence that this behavior mirrors the privacy guarantee of differentially private algorithms under composition. Indeed, the proof of our "advanced" group privacy bound will follow the same strategy as the proof of the so-called advanced composition theorem [11, 10].

**Definition 4.1.** *For random variables $A, B$, define the privacy loss function $\ell_{A,B}(y) = \ln(\Pr[A = y]/\Pr[B = y])$. Define the privacy loss random variable $L_{A,B}$ to be $\ell_{A,B}(y)$ for $y \leftarrow A$.*

Just as in the proof of the advanced composition theorem, we will leverage the fact that the expected privacy loss of any individual local randomizer is $L_{R_i(x_i), R_i(x_i)} = O(\varepsilon^2)$, which is substantially smaller than the worst-case privacy loss of $\varepsilon$. Since each local randomizer is applied independently, the cumulative privacy loss incurred by changing a group of $k$ inputs concentrates to within $O(\sqrt{k}\varepsilon)$ of its expectation $O(k\varepsilon^2)$.

**Theorem 4.2** (Advanced Grouposition for Pure LDP). *Let $x \in X^n$ and $x' \in X^n$ differ in at most $k$ entries for some $1 \leq k \leq n$. Let $\mathcal{A} = (R_1, \ldots, R_n) : X^n \to Y$ be $\varepsilon$-LDP. Then for every $\delta > 0$ and $\varepsilon' = k\varepsilon^2/2 + \varepsilon\sqrt{2k\ln(1/\delta)}$, we have*

$$\Pr_{y \leftarrow \mathcal{A}(x)}\left[\ln \frac{\Pr[\mathcal{A}(x) = y]}{\Pr[\mathcal{A}(x') = y]} > \varepsilon'\right] \leq \delta.$$

*In particular, for every $\delta > 0$ and every set $T \subseteq Y$, we have $\Pr[\mathcal{A}(x) \in T] \leq e^{\varepsilon'} \Pr[\mathcal{A}(x') \in T] + \delta$.*

*Proof.* Without loss of generality, we may assume that $x$ and $x'$ differ in the first $k$ coordinates. Since each randomizer is applied independently, we may write the privacy loss between $\mathcal{A}(x)$ and $\mathcal{A}(x')$ as

$$L_{\mathcal{A}(x), \mathcal{A}(x')} = \sum_{i=1}^{k} L_{R_i(x_i), R_i(x_i')}$$

Since each $R_i$ is $\varepsilon$-differentially private, Proposition 3.3 of [5] implies

$$\mathbb{E}\left[L_{R_i(x_i), R_i(x_i')}\right] \leq \frac{1}{2}\varepsilon^2, \qquad |L_{R_i(x_i), R_i(x_i')}| \leq \varepsilon.$$

Hence, by Hoeffding's inequality, we have that for every $t > 0$,

$$\Pr[L_{R(x), R(x')} > k\varepsilon^2/2 + t] \leq e^{-t^2/2k\varepsilon^2}.$$

Setting $t = \varepsilon\sqrt{2k \ln(1/\delta)}$ completes the proof. $\square$

We remark that it is straightforward to extend Theorem 4.2 to handle $(\varepsilon, \delta)$-LDP algorithms using the same ideas as in the proof of the advanced composition theorem (see the discussion in Section 2.2 of [34]).

**Theorem 4.3** (Advanced Grouposition for Approximate LDP)**.** *Let $x \in X^n$ and $x' \in X^n$ differ in at most $k$ entries for some $1 \leq k \leq n$. Let $\mathcal{A} = (R_1, \ldots, R_n) : X^n \to Y$ be $(\varepsilon, \delta)$-LDP. Then for every $\delta' > 0$, $\varepsilon' = k\varepsilon^2/2 + \varepsilon\sqrt{2k\ln(1/\delta')}$, and every set $T \subseteq Y$, we have $\Pr[\mathcal{A}(x) \in T] \leq e^{\varepsilon'} \Pr[\mathcal{A}(x') \in T] + \delta + k\delta'$.*

Our improved group privacy bound immediately implies a strong bound on the *max-information* of an LDP protocol. The max-information [8] of an algorithm is a measure of how much information it reveals about a randomly chosen input. The motivation for studying max-information comes from its ability to ensure generalization in adaptive data analysis.

**Definition 4.4.** *Let $Z, W$ be jointly distributed random variables. We say that the $\beta$-approximate max-information between $Z$ and $W$, denoted $I_\infty^\beta(Z; W)$, is at most $k$ if for every measurable subset $T$ of the support of $(Z, W)$ with $\Pr[(Z, W) \in T] > \beta$, we have*

$$\ln \frac{\Pr[(Z, W) \in T] - \beta}{\Pr[Z \otimes W \in T]} \leq k.$$

*Here, $Z \otimes W$ denotes the product distribution formed from the marginals $Z$ and $W$. Given an algorithm $\mathcal{A} : X^n \to \mathcal{R}$, we say the $\beta$-approximate max-information of $\mathcal{A}$, denoted $I_\infty^\beta(\mathcal{A}, n) \leq k$ if $I_\infty^\beta(\mathcal{D}; \mathcal{A}(\mathcal{D})) \leq k$ for every distribution $\mathcal{D}$ on $X^n$.*

Dwork et al. [8] showed that $\varepsilon$-DP algorithms have max-information $O(\varepsilon n)$. Moreover, they have $\beta$-approximate max-information $O(\varepsilon^2 n + \varepsilon\sqrt{n \log(1/\beta)})$, but only when the input distribution $\mathcal{D}$ is a product distribution. Subsequent work by Rogers et al. [29] extended this analysis to $(\varepsilon, \delta)$-DP algorithms, for which they showed the restriction to product distributions to be necessary.

Using our group privacy bound for local differential privacy, we show that even under non-product distributions, the max-information of $\varepsilon$-LDP protocols has the same behavior as $\varepsilon$-DP algorithms on product distributions. This provides a distinct advantage in adaptive data analysis, as it means $\varepsilon$-LDP algorithms can be composed with arbitrary low max-information algorithms *in any order* while giving strong generalization guarantees.

**Theorem 4.5.** *Let $\mathcal{A} : X^n \to Y$ be $\varepsilon$-LDP. Then for every $\beta > 0$, we have $I_\infty^\beta(R, n) \leq n\varepsilon^2/2 + \varepsilon\sqrt{2n \ln(1/\beta)}$.*

*Proof.* Let $\mathcal{D}$ be a distribution over $X^n$. Then for every $k > 0$,

$$\Pr_{x \sim \mathcal{D}, y \leftarrow \mathcal{A}(x)} \left[ \ln \frac{\Pr[\mathcal{D} = x, \mathcal{A}(x) = y]}{\Pr[\mathcal{D} = x] \Pr[\mathcal{A}(\mathcal{D}) = y]} > k \right] = \Pr_{x \sim \mathcal{D}, y \leftarrow \mathcal{A}(x)} \left[ \ln \frac{\Pr[\mathcal{A}(x) = y]}{\Pr[\mathcal{A}(\mathcal{D}) = y]} > k \right]$$

$$= \Pr_{x \sim \mathcal{D}, y \leftarrow \mathcal{A}(x)} \left[ \ln \frac{\Pr[\mathcal{A}(x) = y]}{\mathbb{E}_{x' \sim \mathcal{D}} [\Pr[\mathcal{A}(x') = y]]} > k \right]$$

$$\leq \Pr_{x \sim \mathcal{D}, y \leftarrow \mathcal{A}(x)} \left[ \mathbb{E}_{x' \sim \mathcal{D}} \left[ \ln \frac{\Pr[\mathcal{A}(x) = y]}{\Pr[\mathcal{A}(x') = y]} \right] > k \right] \quad \text{by Jensen's inequality}$$

$$\leq \max_{x, x'} \Pr_{y \leftarrow \mathcal{A}(x)} \left[ \ln \frac{\Pr[\mathcal{A}(x) = y]}{\Pr[\mathcal{A}(x') = y]} > k \right].$$

By Theorem 4.2, this quantity as at most $\beta$ for $k = n\varepsilon^2/2 + \varepsilon\sqrt{2n \ln(1/\beta)}$.

The claim now follows from the general fact that, for any pair of random variables $U, V$ over the same sample space,

$$\Pr_{u \sim U} \left[ \ln \frac{\Pr[U = u]}{\Pr[V = u]} > k \right] \leq \beta$$

implies that

$$\sup_{T: \Pr[U \in T] > \beta} \ln \frac{\Pr[U \in T] - \beta}{\Pr[V \in T]} \leq k.$$

□

## 5 Composition for Randomized Response

In this section, we give a direct proof showing that $k$ instantiations of randomized response obey privacy guarantees matching advanced composition, even under pure $\varepsilon$-LDP. Specifically, we exhibit for every $\beta > 0$ an $O(\varepsilon\sqrt{k \ln(1/\beta)})$-LDP algorithm that when run on any given input, yields a distribution which is $\beta$-close in statistical distance to the composition of $k$ instances of randomized response.[5] Prior work of Duchi, Jordan, and Wainwright [7] showed that the *problem* of estimating $k$ binary attributes admits an LDP algorithm with similar guarantees as ours. However, their distribution does not resemble the $k$-fold composition of randomized response. We interpret our new construction as evidence that an advanced composition theorem for pure $\varepsilon$-LDP might hold for more general, and potentially even interactive mechanisms.

Let $M_i : \{0, 1\}^k \to \{0, 1\}$ that performs randomized response on the $i$th bit of its input. On an input $x$, our algorithm first identifies a "good" set $G_x$ of outputs $(y_1, \ldots, y_k) \in \{0, 1\}^k$ which have roughly average likelihood of appearing under the composition $M(x) = (M_1(x), \ldots, M_k(x))$. By concentration arguments, $M(x)$ produces an outcome in $G_x$ with probability at least $1 - \beta$. Our approximate composed algorithm $\tilde{M}(x)$ simply runs $M(x)$, returns the outcome if it is in $G_x$, and returns a uniformly random element outside $G_x$ otherwise. The crux of the privacy argument is to show that elements in this latter case end up being sampled with roughly the same probability as elements in $G_x$.

---

[5]The usual guarantee of advanced composition also includes an additive $O(\varepsilon^2 k)$ term, but our result only applies to the typical setting where $\varepsilon^2 k \leq \varepsilon\sqrt{k}$.

**Theorem 5.1.** *For each $i = 1, \ldots, k$, let $M_i : \{0,1\}^k \to \{0,1\}$ be the $\varepsilon$-differentially private algorithm*

$$M_i(x) = \begin{cases} x_i & \text{w.p. } \frac{e^\varepsilon}{e^\varepsilon+1} \\ 1 - x_i & \text{w.p. } \frac{1}{e^\varepsilon+1}. \end{cases}$$

*Let $0 < \beta < (\varepsilon\sqrt{k}/2(k+1))^{2/3}$ and suppose $\tilde{\varepsilon} = 6\varepsilon\sqrt{k\ln(1/\beta)} \leq 1$. Then there exists an algorithm $\tilde{M} : \{0,1\}^k \to \{0,1\}^k$ such that*

1. *$\tilde{M}$ is $\tilde{\varepsilon}$-differentially private.*

2. *For every $x \in \{0,1\}^k$, there exists an event $E$ with $\Pr[E] \geq 1 - \beta$ such that, conditioned on $E$, the output $\tilde{M}(x)$ is identically distributed to $M(x) = (M_1(x), \ldots, M_k(x))$.*

*Proof.* Let $M(x) = (M_1(x), \ldots, M_k(x))$. For strings $x, y \in \{0,1\}^k$, let $d_H(x,y) = |\{i \in [k] : x_i \neq y_i\}|$ denote the Hamming distance between $x$ and $y$. For each $x \in \{0,1\}^k$, define a "good" spherical shell around $x$ by

$$G_x = \left\{ y \in \{0,1\}^k : \frac{k}{e^\varepsilon+1} - \sqrt{k\ln(2/\beta)/2} \leq d_H(x,y) \leq \frac{k}{e^\varepsilon+1} + \sqrt{k\ln(2/\beta)/2} \right\}$$

It is immediate from Hoeffding's inequality that $\Pr[M(x) \in G_x] \leq \beta$ for all $x \in \{0,1\}^k$.

Our "approximate" composed algorithm is as follows.

---

**Algorithm** Approximate composed algorithm $\tilde{M}(x)$

1. Sample $y = (y_1, \ldots, y_k) \leftarrow M(x)$

2. If $y \in G_x$, output $y$;
   Else, output a random sample from $\{0,1\}^k \setminus G_x$.

---

Since $\Pr[M(x) \in G_x] \geq 1 - \beta$, the accuracy condition (2) in the statement of Theorem 5.1 is immediate. We now show that Algorithm 2 guarantees $\tilde{\varepsilon}$-differential privacy. The main technical lemma that we need to prove this is the following, which says that each $y \notin G_x$ is sampled with probability close to the probability with which any item in $G_x$ is sampled.

**Lemma 5.2.** *For every $x \in \{0,1\}^k$ and $y \notin G_x$,*

$$\Pr[\tilde{M}(x) = y] \in \left[ e^{-3\varepsilon\sqrt{k\ln(1/\beta)}}, e^{3\varepsilon\sqrt{k\ln(1/\beta)}} \right] \cdot \left( \frac{1}{e^\varepsilon+1} \right)^{k/(e^\varepsilon+1)} \left( \frac{e^\varepsilon}{e^\varepsilon+1} \right)^{k/(e^\varepsilon+1)}.$$

*Proof.* Let $U$ be a uniform random variable on $\{0,1\}^k$. If $y \notin G_x$, then

$$\Pr[\tilde{M}(x) = y] = \Pr[M(x) \notin G_x] \cdot \frac{1}{|\{0,1\}^k \setminus G_x|}$$

$$= 2^{-k} \cdot \frac{\Pr[M(x) \notin G_x]}{\Pr[U \notin G_x]}.$$

To analyze this ratio, it suffices by symmetry to consider $x = 0^k$, in which case the set of interest is

$$B = \{0,1\}^k \setminus G_{0^k} = \left\{ y \in \{0,1\}^k : |y| < \frac{k}{e^\varepsilon+1} - \sqrt{k\ln(2/\beta)/2} \text{ or } |y| > \frac{k}{e^\varepsilon+1} + \sqrt{k\ln(2/\beta)/2} \right\}.$$

We further divide up $B$ and consider the set $R$ defined by
$$R = \left\{ y \in \{0,1\}^k : |y| < \frac{k}{e^\varepsilon + 1} - 2\sqrt{k \ln(1/\beta)} \quad \text{or} \quad |y| > \frac{k}{e^\varepsilon + 1} + 2\sqrt{k \ln(1/\beta)} \right\}.$$
Our strategy will be to show that for $y \in B \setminus R$, the ratio $\Pr[M(0^m) = y]/\Pr[U = y]$ is roughly as prescribed. Meanwhile, we will also have $\Pr[M(0^m) \in R] \ll \Pr[U \in B]$ and $\Pr[U \in R] \ll \Pr[U \in B]$, so the contributions of points in $B \setminus R$ will dominate the ratio of interest.

More formally, we make use of the following two claims.

**Claim 5.3.** *For every $y \in B \setminus R$, we have*
$$\frac{\Pr[M(0^m) = y]}{\Pr[U = y]} \in \left[ e^{-2\varepsilon \sqrt{k \ln(1/\beta)}}, e^{2\varepsilon \sqrt{k \ln(1/\beta)}} \right] \cdot 2^k \cdot \left( \frac{1}{e^\varepsilon + 1} \right)^{k/(e^\varepsilon + 1)} \left( \frac{e^\varepsilon}{e^\varepsilon + 1} \right)^{ke^\varepsilon/(e^\varepsilon + 1)}.$$

**Claim 5.4.** *Both of the ratios*
$$\frac{\Pr[M(0^m) \in R]}{\Pr[U \in B]}, \frac{\Pr[U \in R]}{\Pr[U \in B]} \leq (k+1)\beta^{3/2}.$$

These claims are technical and unenlightening, so we defer them to the end of the proof. For now, we show how to put them together to complete the proof of Lemma 5.2. Let
$$C = 2^k \cdot \left( \frac{1}{e^\varepsilon + 1} \right)^{k/(e^\varepsilon + 1)} \left( \frac{e^\varepsilon}{e^\varepsilon + 1} \right)^{ke^\varepsilon/(e^\varepsilon + 1)} \geq 1.$$

$$\begin{aligned}
\Pr[M(0^m) \in B] &= \Pr[M(0^m) \in R] + \Pr[M(0^m) \in B \setminus R] \\
&\leq (k+1)\beta^{3/2} \Pr[U \in B] + e^{2\varepsilon \sqrt{k \ln(2/\beta)}} \cdot C \cdot \Pr[U \in B \setminus R] \\
&\leq (e^{2\varepsilon \sqrt{k \ln(2/\beta)}} \cdot C + (k+1)\beta^{3/2}) \Pr[U \in B] \\
&\leq e^{3\varepsilon \sqrt{k \ln(2/\beta)}} \cdot C \cdot \Pr[U \in B].
\end{aligned}$$

Here, the last inequality follows because $e^{2\varepsilon \sqrt{k \ln(2/\beta)}} + (k+1)\beta^{3/2} \leq e^{3\varepsilon \sqrt{k \ln(2/\beta)}}$ when $(k+1)\beta^{3/2} \leq \tilde{\varepsilon}/12 \leq \varepsilon \sqrt{k \ln(2/\beta)} e^{2\varepsilon \sqrt{k \ln(2/\beta)}}$. Similarly,

$$\begin{aligned}
\Pr[M(0^m) \in B] &\geq \Pr[M(0^m) \in B \setminus R] \\
&\geq e^{-2\varepsilon \sqrt{k \ln(2/\beta)}} \cdot C \cdot \Pr[U \in B \setminus R] \\
&\geq e^{-2\varepsilon \sqrt{k \ln(2/\beta)}} \cdot C \cdot (\Pr[U \in B] - (k+1)\beta^{3/2} \Pr[U \in B]) \\
&= e^{-2\varepsilon \sqrt{k \ln(2/\beta)}} \cdot C \cdot (1 - (k+1)\beta^{3/2}) \cdot \Pr[U \in B] \\
&\geq e^{-3\varepsilon \sqrt{k \ln(2/\beta)}} \cdot C \cdot \Pr[U \in B],
\end{aligned}$$

where the last inequality holds because $(k+1)\beta^{3/2} \leq \tilde{\varepsilon}/12 \leq 1 - e^{-\varepsilon \sqrt{k \ln(2/\beta)}}$. Thus we have
$$2^{-k} \cdot \frac{\Pr[M(0^m) \in B]}{\Pr[U \in B]} \in \left[ e^{-3\varepsilon \sqrt{k \ln(1/\beta)}}, e^{3\varepsilon \sqrt{k \ln(1/\beta)}} \right] \cdot \left( \frac{1}{e^\varepsilon + 1} \right)^{k/(e^\varepsilon + 1)} \left( \frac{e^\varepsilon}{e^\varepsilon + 1} \right)^{ke^\varepsilon/(e^\varepsilon + 1)}.$$
This completes the proof of Lemma 5.2. $\square$

We now use Lemma 5.2 to argue that $\tilde{M}$ is $\tilde{\varepsilon}$-differentially private. Fix two inputs $x, x' \in \{0,1\}^k$. We will show that for every $y \in \{0,1\}^k$, we have $\Pr[\tilde{M}(x) = y] \leq e^{\tilde{\varepsilon}} \Pr[\tilde{M}(x') = y]$. There are four cases to consider.

**Case 1.** $y \in G_x$ and $y \in G_{x'}$. Here

$$\frac{\Pr[\tilde{M}(x) = y]}{\Pr[\tilde{M}(x') = y]} = \frac{\left(\frac{1}{e^\varepsilon+1}\right)^{d_H(x,y)} \left(\frac{e^\varepsilon}{e^\varepsilon+1}\right)^{k-d_H(x,y)}}{\left(\frac{1}{e^\varepsilon+1}\right)^{d_H(x',y)} \left(\frac{e^\varepsilon}{e^\varepsilon+1}\right)^{k-d_H(x',y)}}$$

$$= \left(\frac{1}{e^\varepsilon+1}\right)^{d_H(x,y)-d_H(x',y)} \left(\frac{e^\varepsilon}{e^\varepsilon+1}\right)^{d_H(x',y)-d_H(x,y)}$$

$$\leq \left(\frac{1}{e^\varepsilon+1}\right)^{-2\sqrt{k\ln(2/\beta)/2}} \left(\frac{e^\varepsilon}{e^\varepsilon+1}\right)^{2\sqrt{k\ln(2/\beta)/2}}$$

$$= e^{\varepsilon\sqrt{2k\ln(2/\beta)}} \leq e^{\tilde{\varepsilon}}.$$

**Case 2.** $y \in G_x$ and $y \notin G_{x'}$. Here

$$\frac{\Pr[\tilde{M}(x) = y]}{\Pr[\tilde{M}(x') = y]} \leq \frac{\left(\frac{1}{e^\varepsilon+1}\right)^{d_H(x,y)} \left(\frac{e^\varepsilon}{e^\varepsilon+1}\right)^{k-d_H(x,y)}}{e^{-3\varepsilon\sqrt{k\ln(2/\beta)}} \cdot \left(\frac{1}{e^\varepsilon+1}\right)^{k/(e^\varepsilon+1)} \left(\frac{e^\varepsilon}{e^\varepsilon+1}\right)^{ke^\varepsilon/(e^\varepsilon+1)}}$$

$$= e^{3\varepsilon\sqrt{k\ln(2/\beta)}} \cdot \left(\frac{1}{e^\varepsilon+1}\right)^{d_H(x,y)-k/(e^\varepsilon+1)} \left(\frac{e^\varepsilon}{e^\varepsilon+1}\right)^{k/(e^\varepsilon+1)-d_H(x,y)}$$

$$\leq e^{3\varepsilon\sqrt{k\ln(2/\beta)}} \cdot \left(\frac{1}{e^\varepsilon+1}\right)^{-\sqrt{k\ln(2/\beta)/2}} \left(\frac{e^\varepsilon}{e^\varepsilon+1}\right)^{\sqrt{k\ln(2/\beta)/2}}$$

$$\leq e^{(3+\sqrt{2})\varepsilon\sqrt{k\ln(2/\beta)}} \leq e^{\tilde{\varepsilon}}.$$

**Case 3.** $y \notin G_x$ and $y \in G_{x'}$. Here

$$\frac{\Pr[\tilde{M}(x) = y]}{\Pr[\tilde{M}(x') = y]} \leq \frac{e^{3\varepsilon\sqrt{k\ln(2/\beta)}} \cdot \left(\frac{1}{e^\varepsilon+1}\right)^{k/(e^\varepsilon+1)} \left(\frac{e^\varepsilon}{e^\varepsilon+1}\right)^{ke^\varepsilon/(e^\varepsilon+1)}}{\left(\frac{1}{e^\varepsilon+1}\right)^{d_H(x',y)} \left(\frac{e^\varepsilon}{e^\varepsilon+1}\right)^{k-d_H(x',y)}}$$

$$= e^{3\varepsilon\sqrt{k\ln(2/\beta)}} \cdot \left(\frac{1}{e^\varepsilon+1}\right)^{k/(e^\varepsilon+1)-d_H(x',y)} \left(\frac{e^\varepsilon}{e^\varepsilon+1}\right)^{d_H(x',y)-k/(e^\varepsilon+1)}$$

$$\leq e^{3\varepsilon\sqrt{k\ln(2/\beta)}} \cdot \left(\frac{1}{e^\varepsilon+1}\right)^{-\sqrt{k\ln(2/\beta)/2}} \left(\frac{e^\varepsilon}{e^\varepsilon+1}\right)^{\sqrt{k\ln(2/\beta)/2}}$$

$$= e^{(3+\sqrt{2})\varepsilon\sqrt{k\ln(2/\beta)}} \leq e^{\tilde{\varepsilon}}.$$

**Case 4.** $y \notin G_x$ and $y \notin G_{x'}$. Here

$$\frac{\Pr[\tilde{M}(x) = y]}{\Pr[\tilde{M}(x') = y]} \leq \frac{e^{3\varepsilon\sqrt{k\ln(2/\beta)}} \cdot \left(\frac{1}{e^\varepsilon+1}\right)^{k/(e^\varepsilon+1)} \left(\frac{e^\varepsilon}{e^\varepsilon+1}\right)^{ke^\varepsilon/(e^\varepsilon+1)}}{e^{-3\varepsilon\sqrt{k\ln(2/\beta)}} \cdot \left(\frac{1}{e^\varepsilon+1}\right)^{k/(e^\varepsilon+1)} \left(\frac{e^\varepsilon}{e^\varepsilon+1}\right)^{ke^\varepsilon/(e^\varepsilon+1)}}$$

$$= e^{6\varepsilon\sqrt{k\ln(2/\beta)}} \leq e^{\tilde{\varepsilon}}.$$

In all cases, we thus have $\Pr[\tilde{M}(x) = y] \le e^{\tilde{\varepsilon}} \Pr[\tilde{M}(x') = y]$. Modulo the deferred claims, this completes the proof of Theorem 5.1. $\square$

*Proof of Claim 5.3.* Let $y \in B \setminus R$. We first consider the case where

$$\frac{k}{e^\varepsilon + 1} - 2\sqrt{k \ln(2/\beta)} \le |y| \le \frac{k}{e^\varepsilon + 1} - \sqrt{k \ln(2/\beta)/2}.$$

Here, we have

$$\frac{\Pr[M(0^k) = y]}{\Pr[U = y]} = 2^k \left(\frac{1}{e^\varepsilon + 1}\right)^{|y|} \left(\frac{e^\varepsilon}{e^\varepsilon + 1}\right)^{k-|y|}$$

$$= \exp\left(\varepsilon \cdot \left(\frac{k}{e^\varepsilon + 1} - |y|\right)\right) \cdot 2^k \cdot \left(\frac{1}{e^\varepsilon + 1}\right)^{k/(e^\varepsilon+1)} \left(\frac{e^\varepsilon}{e^\varepsilon + 1}\right)^{ke^\varepsilon/(e^\varepsilon+1)}$$

$$\in \left[1, e^{2\varepsilon\sqrt{k\ln(2/\beta)}}\right] \cdot 2^k \cdot \left(\frac{1}{e^\varepsilon + 1}\right)^{k/(e^\varepsilon+1)} \left(\frac{e^\varepsilon}{e^\varepsilon + 1}\right)^{ke^\varepsilon/(e^\varepsilon+1)}.$$

An identical argument shows that when

$$\frac{k}{e^\varepsilon + 1} + \sqrt{k \ln(2/\beta)/2} \le |y| \le \frac{k}{e^\varepsilon + 1} + 2\sqrt{k \ln(2/\beta)},$$

we have the relationship

$$\frac{\Pr[M(0^k) = y]}{\Pr[U = y]} \in \left[e^{-2\varepsilon\sqrt{k\ln(2/\beta)}}, 1\right] \cdot 2^k \cdot \left(\frac{1}{e^\varepsilon + 1}\right)^{k/(e^\varepsilon+1)} \left(\frac{e^\varepsilon}{e^\varepsilon + 1}\right)^{ke^\varepsilon/(e^\varepsilon+1)}.$$

$\square$

Before proving Claim 5.4, we state and prove the following anti-concentration result for uniform strings.

**Lemma 5.5.** *Let $U$ be uniformly distributed on $\{0,1\}^k$. Then for every $t \in [0, \sqrt{k}/2]$,*

$$\Pr\left[|U| \ge \frac{k}{2} + t\sqrt{k}\right] \ge \frac{\exp(-3t^2)}{k+1}.$$

*Proof.* Using the probability mass function of the binomial distribution, we calculate

$$\Pr\left[|U| \ge \frac{k}{2} + t\sqrt{k}\right] = 2^{-k} \sum_{j=0}^{k/2 - \lfloor t\sqrt{k} \rfloor} \binom{k}{j} \ge 2^{-k} \binom{k}{k/2 - \lfloor t\sqrt{k} \rfloor}.$$

Using the consequence of Stirling's approximation that $\binom{k}{j} \ge 2^{k \cdot H(j/k)}/(k+1)$ where $H$ is the binary entropy function, this is at least

$$2^{-k} \cdot \frac{2^{k \cdot H(1/2 - t/\sqrt{k})}}{k+1}.$$

Now we use the fact that $H(1/2 - \eta) \ge 1 - 4\eta^2$ to lower bound this by

$$\frac{2^{-4t^2}}{k+1}.$$

$\square$

*Proof of Claim 5.4.* By the lower bound of Lemma 5.5 on the tail bound of the uniform distribution,

$$\Pr[U \in B] \geq \Pr\left[|U| > \frac{k}{e^\varepsilon + 1} + \sqrt{k \ln(2/\beta)/2}\right]$$

$$= \Pr\left[|U| > \frac{k}{2} + \left(\sqrt{\ln(2/\beta)/2} - \frac{(e^\varepsilon - 1)\sqrt{k}}{2(e^\varepsilon + 1)}\right)\sqrt{k}\right]$$

$$\geq \frac{1}{k+1} \exp\left(-3 \cdot \left(\sqrt{\ln(2/\beta)/2} - \frac{(e^\varepsilon - 1)\sqrt{k}}{2(e^\varepsilon + 1)}\right)^2\right).$$

On the other hand, by Hoeffding's inequality,

$$\Pr[M(0^m) \in R] \leq 2 \cdot \exp\left(-2 \cdot (2\sqrt{\ln(2/\beta)})^2\right),$$

and

$$\Pr[U \in R] \leq 2 \cdot \exp\left(-2 \cdot \left(2\sqrt{\ln(2/\beta)} - \frac{(e^\varepsilon - 1)\sqrt{k}}{2(e^\varepsilon + 1)}\right)^2\right).$$

This allows us to conclude that both of the ratios $\Pr[M(0^m) \in R]/\Pr[U \in B]$ and $\Pr[U \in R]/\Pr[U \in B]$ are at most

$$2(k+1) \cdot \exp\left(-\frac{13}{2}\ln(2/\beta) + (8 - 3\sqrt{2})\sqrt{\ln(2/\beta)} \cdot \frac{(e^\varepsilon - 1)\sqrt{k}}{2(e^\varepsilon + 1)} + \left(\frac{(e^\varepsilon - 1)\sqrt{k}}{2(e^\varepsilon + 1)}\right)^2\right).$$

Using the fact that $(e^\varepsilon - 1)\sqrt{k}/2(e^\varepsilon + 1) \leq \sqrt{\ln(2/\beta)}$, this is at most

$$2(k+1) \cdot \exp\left(-\frac{3}{2}\ln(2/\beta)\right) \leq (k+1) \cdot \beta^{3/2}.$$

□

# 6 From Approximate-Privacy to Pure-Privacy: A Generic Transformation with Short Reports

In this section we present our generic transformation from any (non-interactive) $(\varepsilon, \delta)$-LDP protocol into an $O(\varepsilon)$-LDP protocol with essentially the same utility guarantees. Our transformation is based on *rejection sampling*, an idea that first appeared in the context of differential privacy in work of Kasiviswanathan et al. [19], where it was used to show an equivalence between the local model (under pure-privacy) and SQ learning. Rejection sampling was also used by Bassily and Smith [4] to show that almost any (non-interactive) $\varepsilon$-LDP protocol can be transformed into a protocol where the communication per user is only 1 bit (at the expense of increasing the shared randomness in the protocol). Our transformation is obtained from the transformation of Bassily and Smith with the following two modifications:

1. The transformation of Bassily and Smith applies to any "sample resilient" $\varepsilon$-LDP protocol. Informally, a sample resilient protocol is one whose outcome on any (distributed) database $S$ is well-approximated by its outcome on a random subset of (roughly) half of the users in $S$. While we do not know of any protocols which are not sample resilient, we would like to obtain a more general result which avoids this restriction. We show that this restriction can be removed at the expense of increasing per-user communication to $O(\log \log n)$, where $n$ is the number of users.

2. We generalize the transformation so that it also holds for any $(\varepsilon, \delta)$-LDP protocol. Somewhat surprisingly, instead of obtaining an $(\varepsilon, \delta)$-LDP protocol with short user reports, we show that, when done carefully, the resulting protocol actually satisfies pure $O(\varepsilon)$-privacy.

Our generic transformation is given in algorithm `GenProt`.

---

**Algorithm `GenProt`**

---

**Inputs:** User's inputs $\{x_i \in X : i \in [n]\}$, privacy parameter $\varepsilon$, and parameter $T \in \mathbb{N}$.

**Algorithms used:** An $(\varepsilon, \delta)$-LDP protocol $\mathcal{M}$ with local randomizers $\mathcal{A}_i$, $i \in [n]$.

1. For every $(i, t) \in [n] \times [T]$ generate independent public string: $y_{i,t} \leftarrow \mathcal{A}_i(\bot)$.

2. For user $i = 1$ to $n$ do

    (a) For every $t \in [T]$ compute $p_{i,t} = \frac{1}{2} \frac{\Pr[\mathcal{A}_i(x_i) = y_{i,t}]}{\Pr[\mathcal{A}_i(\bot) = y_{i,t}]}$.

    (b) For every $t \in [T]$, if $p_{i,t} \notin \left[\frac{e^{-2\varepsilon}}{2}, \frac{e^{2\varepsilon}}{2}\right]$ then set $p_{i,t} = \frac{1}{2}$.

    (c) For every $t \in [T]$ sample a bit $b_{i,t}$ from Bernoulli($p_{i,t}$).

    (d) Denote $H_i = \{t \in [T] : b_{i,t} = 1\}$.

    (e) If $H_i = \emptyset$ then set $H_i = [T]$.

    (f) Sample $g_i \in H_i$ uniformly, and send $g_i$ to the server.

3. Run $\mathcal{M}(y_{1,g_1}, \ldots, y_{n,g_n})$.

---

**Theorem 6.1.** *Let $\varepsilon \leq 1/4$, and let*

$$5 \ln(1/\varepsilon) \leq T \leq \frac{1 - e^{-\varepsilon}}{4\delta e^\varepsilon n}.$$

*Then Algorithm `GenProt` satisfies $10\varepsilon$-LDP. Moreover, for every database $S = (x_1, \ldots, x_n)$, the total variation distance between `GenProt`(S) and $\mathcal{M}(S)$ is at most*

$$\sup_T |\Pr[\textit{GenProt}(S) \in T] - \Pr[\mathcal{M}(S) \in T]| \leq n \left(\left(\frac{1}{2} + \varepsilon\right)^T + \frac{6T\delta e^\varepsilon}{1 - e^{-\varepsilon}}\right).$$

To interpret the guarantee of this theorem, let $\beta > 0$ be a parameter, and suppose $\delta \leq \varepsilon\beta/48n \ln(2n/\beta)$. Then setting $T = 2\ln(2n/\beta)$ makes this total variation distance at most $\beta$. With this setting of parameters, observe that each user must send $O(\log \log n)$ bits to the server.

We prove the privacy and utility guarantees of Theorem 6.1 individually.

**Lemma 6.2.** *Let $\varepsilon \leq 1/4$ and let $T \geq 5\ln(1/\varepsilon)$. Then Algorithm* `GenProt` *satisfies $10\varepsilon$-LDP.*

*Proof.* Fix the public randomness $y_{i,t}$ for $(i,t) \in [n] \times [T]$. Fix a user $i \in [n]$, and let $\mathcal{Q}_i$ denote the output of user $i$ in the protocol (from step 2f) after fixing the public randomness. Let $x_i, x_i' \in X$ be two possible inputs, and let $g \in [T]$ be a possible output of $\mathcal{Q}_i$. Our goal is to show that

$$\Pr[\mathcal{Q}_i(x_i) = g] \leq e^{10\varepsilon} \cdot \Pr[\mathcal{Q}_i(x_i') = g].$$

We calculate:

$$\Pr[\mathcal{Q}_i(x_i) = g] = \Pr[b_{i,g} = 1] \cdot \Pr[\mathcal{Q}_i(x_i) = g | b_{i,g} = 1] + \Pr[b_{i,g} = 0] \cdot \Pr[\mathcal{Q}_i(x_i) = g | b_{i,g} = 0]. \quad (5)$$

We now analyze each of these two terms. First observe that if $b_{i,g} = 0$, then for $g$ to be a possible output it must be that $H_i = \emptyset$ before step 2e, and that the output is chosen uniformly from $H_i = [T]$. So,

$$\Pr[\mathcal{Q}_i(x_i) = g | b_{i,g} = 0] = \frac{1}{T} \cdot \Pr[\forall t \neq g \text{ we have } b_{i,t} = 0] \leq \frac{1}{T}\left(\frac{e^{2\varepsilon}}{2}\right)^{T-1}.$$

Next note that if $b_{i,g} = 1$ then we have that $H_i \neq \emptyset$ before step 2e, and that the output is chosen uniformly from $H_i$. So,

$$\begin{aligned}\Pr[\mathcal{Q}_i(x_i) = g | b_{i,g} = 1] &= \sum_{k=1}^{T} \Pr[|H_i| = k | b_{i,g} = 1] \cdot \frac{1}{k} = \sum_{k=1}^{T} \Pr[|H_i \setminus \{g\}| = k-1] \cdot \frac{1}{k} \\ &= \sum_{k=0}^{T-1} \Pr[|H_i \setminus \{g\}| = k] \cdot \frac{1}{k+1} = \mathbb{E}\left[\frac{1}{W+1}\right],\end{aligned} \quad (6)$$

where $W \triangleq |H_i \setminus \{g\}| = \sum_{t \neq g} b_{i,t}$ is a sum of independent Bernoulli random variables, with different expectations $p_{i,t} \in \left[\frac{e^{-2\varepsilon}}{2}, \frac{e^{2\varepsilon}}{2}\right]$ (this is called the *Poisson binomial distribution*). We now want to relate the random variable $W$ to the binomial distribution, specifically, to a random variable $\hat{W}$ defined as the sum of $(T-1)$ i.i.d. Bernoullis with expectations $\hat{p} = \frac{e^{-2\varepsilon}}{2}$. This is useful as the binomial distribution is much easier to analyze than the Poisson binomial distribution.

Our current task is to show that $\mathbb{E}[1/(W+1)] \leq \mathbb{E}[1/(\hat{W}+1)]$. Intuitively, observe that $W$ is the sum of $T-1$ Bernoullis: $W = \sum_{t \neq q} b_{i,t}$, and recall that we defined $\hat{W}$ by *decreasing* the expectations of each of these Bernoulli random variables. Hence, intuitively we should get that $\mathbb{E}[1/(W+1)] \leq \mathbb{E}[1/(\hat{W}+1)]$. We now make this argument formal.

**Claim 6.3.** *Let $X \sim \text{Bernoulli}(p)$ and $\hat{X} \sim \text{Bernoulli}(\hat{p})$ where $p \geq \hat{p}$. Let $Y \geq 1$ be independent of $X, \hat{X}$. Then $\mathbb{E}\left[\frac{1}{X+Y}\right] \leq \mathbb{E}\left[\frac{1}{\hat{X}+Y}\right]$.*

*Proof.*

$$\begin{aligned}\mathbb{E}\left[\frac{1}{X+Y}\right] &= \int_0^\infty \Pr\left[\frac{1}{X+Y} \geq z\right] dz = \int_0^\infty \sum_y \Pr[Y=y] \Pr\left[\frac{1}{X+y} \geq z\right] dz \\ &\leq \int_0^\infty \sum_y \Pr[Y=y] \Pr\left[\frac{1}{\hat{X}+y} \geq z\right] dz = \mathbb{E}\left[\frac{1}{\hat{X}+Y}\right].\end{aligned}$$

□

Using Claim 6.3 we can replace, one by one, every Bernoulli($p$) variable in the sum of $W$ with a Bernoulli($\hat{p}$) variable, without decreasing the expectation $\mathbb{E}[1/(W+1)]$. So,

$$
\begin{aligned}
(6) \leq \mathbb{E}[1/(\hat{W}+1)] &= \sum_{k=0}^{T-1} \frac{1}{k+1}\binom{T-1}{k} \cdot \hat{p}^k \cdot (1-\hat{p})^{T-1-k} \\
&= \sum_{k=0}^{T-1} \frac{1}{k+1} \frac{(T-1)!}{k!(T-1-k)!} \cdot \hat{p}^k \cdot (1-\hat{p})^{T-1-k} \\
&= \sum_{k=0}^{T-1} \frac{1}{T} \frac{T!}{(k+1)!(T-1-k)!} \cdot \hat{p}^k \cdot (1-\hat{p})^{T-1-k} \\
&= \frac{1}{T} \sum_{k=0}^{T-1} \binom{T}{k+1} \cdot \hat{p}^k \cdot (1-\hat{p})^{T-1-k} \\
&= \frac{1}{T} \sum_{\ell=1}^{T} \binom{T}{\ell} \cdot \hat{p}^{\ell-1} \cdot (1-\hat{p})^{T-\ell} \\
&= \frac{1}{T\hat{p}} \sum_{\ell=1}^{T} \binom{T}{\ell} \cdot \hat{p}^{\ell} \cdot (1-\hat{p})^{T-\ell} \\
&= \frac{1}{T\hat{p}} \left[ -(1-\hat{p})^T + \sum_{\ell=0}^{T} \binom{T}{\ell} \cdot \hat{p}^{\ell} \cdot (1-\hat{p})^{T-\ell} \right] \\
&= \frac{1}{T\hat{p}} \left[ -(1-\hat{p})^T + 1 \right] = \frac{2}{Te^{-2\varepsilon}}\left[1 - \left(1 - \frac{e^{-2\varepsilon}}{2}\right)^T\right].
\end{aligned}
$$

Going back to Equation (5) we have that

$$
\begin{aligned}
\Pr[\mathcal{Q}_i(x_i) = g] &= \Pr[b_{i,g}=1] \cdot \Pr[\mathcal{Q}_i(x_i)=g|b_{i,g}=1] + \Pr[b_{i,g}=0] \cdot \Pr[\mathcal{Q}_i(x_i)=g|b_{i,g}=0] \\
&\leq \frac{e^{2\varepsilon}}{2} \cdot \frac{2}{Te^{-2\varepsilon}}\left[1 - \left(1 - \frac{e^{-2\varepsilon}}{2}\right)^T\right] + \frac{e^{2\varepsilon}}{2} \cdot \frac{1}{T}\left(\frac{e^{2\varepsilon}}{2}\right)^{T-1} \\
&= \frac{1}{T}\left[e^{4\varepsilon} - e^{4\varepsilon}\left(1 - \frac{e^{-2\varepsilon}}{2}\right)^T + \left(\frac{e^{2\varepsilon}}{2}\right)^T\right] \leq \frac{1}{T}\left[e^{4\varepsilon} + \left(\frac{e^{2\varepsilon}}{2}\right)^T\right]
\end{aligned}
$$

An identical analysis shows that

$$
\Pr[\mathcal{Q}_i(x'_i) = g] \geq \frac{1}{T}\left[e^{-4\varepsilon} - e^{-4\varepsilon}\left(1 - \frac{e^{2\varepsilon}}{2}\right)^T\right].
$$

So,

$$
\frac{\Pr[\mathcal{Q}_i(x_i) = g]}{\Pr[\mathcal{Q}_i(x'_i) = g]} \leq \frac{e^{4\varepsilon} + \left(\frac{e^{2\varepsilon}}{2}\right)^T}{e^{-4\varepsilon} - e^{-4\varepsilon}\left(1 - \frac{e^{2\varepsilon}}{2}\right)^T},
$$

which is at most $e^{10\varepsilon}$ for $\varepsilon \leq 1/4$ and $T \geq 5\ln(1/\varepsilon)$. $\square$

**Lemma 6.4.** *Let $T \leq \frac{1-e^{-\varepsilon}}{4\delta e^{\varepsilon} n}$. Then any event that occurs with probability $p$ under the original protocol happens with probability at most*

$$p + n\left(\left(\frac{1}{2} + \varepsilon\right)^T + \frac{6T\delta e^{\varepsilon}}{1 - e^{-\varepsilon}}\right)$$

*in the execution of* `GenProt`.

*Proof.* Fix user inputs $x_1, \ldots, x_n$. For $(i,t) \in [n] \times [T]$ we denote by $Y_{i,t}$ the random variable taking value $y_{i,t}$ in line 1 of `GenProt`. We also denote $Y_i = Y_{i,G_i}$, where $G_i$ is a random variable taking value $g_i$ in line 2f of `GenProt`. With these notations, our goal is to relate the following two random variables:

$$\mathcal{M}(\mathcal{A}_1(x_1), \ldots, \mathcal{A}_n(x_n)) \quad \text{and} \quad \mathcal{M}(Y_1, \ldots, Y_n).$$

To that end, we now define the following events:

> **Event $E$:** For every $i \in [n]$ we have that $H_i \neq \emptyset$ before step 2e.
>
> **Event $E_i$:** We have that $H_i \neq \emptyset$ before step 2e.

We show that $E$ happens with high probability. Fix $i \in [n]$. For $\varepsilon \leq \frac{1}{4}$ and $\beta = n(1/2 + \varepsilon)^T$ we have

$$\Pr[H_i = \emptyset] \leq \left(1 - \frac{e^{-2\varepsilon}}{2}\right)^T \leq \left(\frac{1}{2} + \varepsilon\right)^T \leq \frac{\beta}{n}.$$

Hence, using the union bound, we get that $\Pr[E] \geq 1 - \beta$.

For every $i \in [n]$ define the set of all "good" random strings $\texttt{Good}_i = \left\{y : \frac{\Pr[\mathcal{A}_i(x_i) = y]}{\Pr[\mathcal{A}_i(\bot) = y]} \in e^{\pm 2\varepsilon}\right\}$.

> **Event Priv:** For every $i \in [n]$ and for every $t \in [T]$ we have that $Y_{i,t} \in \texttt{Good}_i$.
>
> **Event Priv$_i$:** For every $t \in [T]$ we have that $Y_{i,t} \in \texttt{Good}_i$.
>
> **Event Priv$_{i,t}$:** We have that $Y_{i,t} \in \texttt{Good}_i$.

To analyze Event Priv we recall the following useful consequence of the definition of differential privacy:

**Observation 6.5** ([20]). *Let $M: X^n \to Y$ be $(\varepsilon, \delta)$-differentially private, and fix neighboring databases $S, S' \in X^n$. Then,*

$$\Pr\left[M(S) = y \text{ s.t. } \frac{\Pr[M(S) = y]}{\Pr[M(S') = y]} \notin e^{\pm 2\varepsilon}\right] \leq \frac{2\delta e^{\varepsilon}}{1 - e^{-\varepsilon}} = O(\delta/\varepsilon).$$

By Observation 6.5 and a union bound for every $(i,t) \in [n] \times [T]$ we have that $\Pr[\text{Priv}] \geq 1 - \frac{2nT\delta e^{\varepsilon}}{1 - e^{-\varepsilon}}$.

For $(i,t) \in [n] \times [T]$ let $B_{i,t}$ be a random variable taking value $b_{i,t}$ in line 2c of `GenProt`. Fix $i \in [n]$ and fix $y \in \texttt{Good}_i$. We calculate:

$$
\begin{aligned}
\Pr[Y_i = y | E, \text{Priv}] &= \Pr[Y_i = y | E_i, \text{Priv}_i] \\
&= \sum_{t \in [T]} \Pr[G_i = t | E_i, \text{Priv}_i] \cdot \Pr[Y_i = y | G_i = t, E_i, \text{Priv}_i] \\
&= \sum_{t \in [T]} \Pr[G_i = t | E_i, \text{Priv}_i] \cdot \Pr[Y_{i,t} = y | G_i = t, B_{i,t} = 1, \text{Priv}_i] \\
&= \sum_{t \in [T]} \Pr[G_i = t | E_i, \text{Priv}_i] \cdot \frac{\Pr[Y_{i,t} = y, G_i = t, B_{i,t} = 1, \text{Priv}_i]}{\Pr[G_i = t, B_{i,t} = 1, \text{Priv}_i]} \\
&= \sum_{t \in [T]} \Pr[G_i = t | E_i, \text{Priv}_i] \cdot \frac{\Pr[G_i = t | Y_{i,t} = y, B_{i,t} = 1, \text{Priv}_i]}{\Pr[G_i = t | B_{i,t} = 1, \text{Priv}_i]} \cdot \Pr[Y_{i,t} = y | B_{i,t} = 1, \text{Priv}_i]
\end{aligned}
$$
(7)

Now recall that $G_i$ is a (random) function of $B_{i,1}, \ldots, B_{i,T}$, and observe that given $B_{i,t} = 1$, we have that $G_i$ is independent of $Y_{i,t}$. Hence,

$$
\begin{aligned}
(7) &= \sum_{t \in [T]} \Pr[G_i = t | E_i, \text{Priv}_i] \cdot \Pr[Y_{i,t} = y | B_{i,t} = 1, \text{Priv}_i] \\
&= \sum_{t \in [T]} \Pr[G_i = t | E_i, \text{Priv}_i] \cdot \Pr[Y_{i,1} = y | B_{i,1} = 1, \text{Priv}_{i,1}] \\
&= \Pr[Y_{i,1} = y | B_{i,1} = 1, \text{Priv}_{i,1}] \cdot \sum_{t \in [T]} \Pr[G_i = t | E_i, \text{Priv}_i] \\
&= \Pr[Y_{i,1} = y | B_{i,1} = 1, \text{Priv}_{i,1}] \\
&= \frac{\Pr[Y_{i,1} = y, B_{i,1} = 1, \text{Priv}_{i,1}]}{\Pr[B_{i,1} = 1, \text{Priv}_{i,1}]}
\end{aligned}
$$
(8)

Recall that we fixed $y$ s.t. $y \in \text{Good}_i$. Hence,

$$
\begin{aligned}
(8) &= \frac{\Pr[Y_{i,1} = y, B_{i,1} = 1]}{\Pr[B_{i,1} = 1, \text{Priv}_{i,1}]} \\
&= \frac{\Pr[Y_{i,1} = y] \cdot \Pr[B_{i,1} = 1 | Y_{i,1} = y]}{\Pr[B_{i,1} = 1, \text{Priv}_{i,1}]} \\
&= \frac{\Pr[Y_{i,1} = y] \cdot \frac{1}{2} \frac{\Pr[\mathcal{A}_i(x_i) = y]}{\Pr[Y_{i,1} = y]}}{\Pr[B_{i,1} = 1, \text{Priv}_{i,1}]} \\
&= \frac{\frac{1}{2} \Pr[\mathcal{A}_i(x_i) = y]}{\Pr[B_{i,1} = 1, \text{Priv}_{i,1}]}
\end{aligned}
$$
(9)

We now analyze $\Pr[B_{i,1} = 1, \text{Priv}_{i,1}]$:

$$\Pr[B_{i,1} = 1, \text{Priv}_{i,1}] = \sum_{y \in \text{Good}_i} \Pr[Y_{i,1} = y] \cdot \Pr[B_{i,1} = 1 | Y_{i,1} = y]$$

$$= \sum_{y \in \text{Good}_i} \Pr[Y_{i,1} = y] \cdot \frac{1}{2} \frac{\Pr[\mathcal{A}_i(x_i) = y]}{\Pr[Y_{i,1} = y]}$$

$$= \frac{1}{2} \sum_{y \in \text{Good}_i} \Pr[\mathcal{A}_i(x_i) = y] = \frac{1}{2} \Pr[\mathcal{A}_i(x_i) \in \text{Good}_i].$$

Hence, as $y \in \text{Good}_i$, we have

$$(9) \quad = \quad \frac{\Pr[\mathcal{A}_i(x_i) = y]}{\Pr[\mathcal{A}_i(x_i) \in \text{Good}_i]} = \Pr[\mathcal{A}_i(x_i) = y \mid \mathcal{A}_i(x_i) \in \text{Good}_i].$$

That is, we have established that for every $i \in [n]$ and for every $y \in \text{Good}_i$ the following holds:

$$\Pr[Y_i = y | E, \text{Priv}] = \Pr[\mathcal{A}_i(x_i) = y \mid \mathcal{A}_i(x_i) \in \text{Good}_i].$$

Let us denote $\text{Good} = \text{Good}_1 \times \text{Good}_2 \times \ldots \times \text{Good}_n$. For simplicity we now assume that the post-processing procedure $\mathcal{M}$ is deterministic. For a subset $F$ of possible outputs of $\mathcal{M}$ we denote

$$F_Y = \{(y_1, \ldots, y_n) : \mathcal{M}(y_1, \ldots, y_n) \in F\}.$$

With this notation we have that

$$\Pr[\mathcal{M}(Y_1, \ldots, Y_n) \in F] \leq \Pr[\overline{E}] + \Pr[\overline{\text{Priv}}] + \Pr[\mathcal{M}(Y_1, \ldots, Y_n) \in F | E, \text{Priv}]$$

$$\leq \beta + \frac{2nT\delta e^\varepsilon}{1 - e^{-\varepsilon}} + \Pr[\mathcal{M}(Y_1, \ldots, Y_n) \in F | E, \text{Priv}]$$

$$= \beta + \frac{2nT\delta e^\varepsilon}{1 - e^{-\varepsilon}} + \sum_{(y_1,\ldots,y_n) \in F_Y} \Pr[Y_1 = y_1, \ldots, Y_n = y_n | E, \text{Priv}]$$

$$= \beta + \frac{2nT\delta e^\varepsilon}{1 - e^{-\varepsilon}} + \sum_{(y_1,\ldots,y_n) \in F_Y} \prod_{i=1}^n \Pr[Y_i = y_i | E, \text{Priv}]$$

$$= \beta + \frac{2nT\delta e^\varepsilon}{1 - e^{-\varepsilon}} + \sum_{(y_1,\ldots,y_n) \in F_Y \cap \text{Good}} \prod_{i=1}^n \Pr[Y_i = y_i | E, \text{Priv}]$$

$$= \beta + \frac{2nT\delta e^\varepsilon}{1 - e^{-\varepsilon}} + \sum_{(y_1,\ldots,y_n) \in F_Y \cap \text{Good}} \prod_{i=1}^n \Pr[\mathcal{A}_i(x_i) = y_i \mid \mathcal{A}_i(x_i) \in \text{Good}_i]$$

$$= \beta + \frac{2nT\delta e^\varepsilon}{1 - e^{-\varepsilon}} + \sum_{(y_1,\ldots,y_n) \in F_Y} \prod_{i=1}^n \Pr[\mathcal{A}_i(x_i) = y_i \mid \mathcal{A}_i(x_i) \in \text{Good}_i] \quad (10)$$

Let us use the shorthand $\left\{\vec{\mathcal{A}}(\vec{x}) \in \text{Good}\right\}$ to denote $\{\mathcal{A}_1(x_1) \in \text{Good}_1, \ldots, \mathcal{A}_n(x_n) \in \text{Good}_n\}$. So,

$$
\begin{align}
(11) \quad &= \beta + \frac{2nT\delta e^\varepsilon}{1-e^{-\varepsilon}} + \sum_{(y_1,\ldots,y_n)\in F_Y} \Pr[\mathcal{A}_1(x_1)=y_1,\ldots,\mathcal{A}_n(x_n)=y_n | \vec{\mathcal{A}}(\vec{x}) \in \texttt{Good}] \\
&= \beta + \frac{2nT\delta e^\varepsilon}{1-e^{-\varepsilon}} + \Pr[\mathcal{M}(\mathcal{A}_1(x_1),\ldots,\mathcal{A}_n(x_n)) \in F | \vec{\mathcal{A}}(\vec{x}) \in \texttt{Good}] \\
&\leq \beta + \frac{2nT\delta e^\varepsilon}{1-e^{-\varepsilon}} + \frac{1}{\Pr[\vec{\mathcal{A}}(\vec{x}) \in \texttt{Good}]} \cdot \Pr[\mathcal{M}(\mathcal{A}_1(x_1),\ldots,\mathcal{A}_n(x_n)) \in F] \tag{11}
\end{align}
$$

Similarly to the analysis of Event Priv, by the privacy properties of the $\mathcal{A}_i$'s we have that $\Pr[\vec{\mathcal{A}}(\vec{x}) \in \texttt{Good}] \geq 1 - \frac{2nT\delta e^\varepsilon}{1-e^{-\varepsilon}}$. Using the fact that $\frac{1}{1-\zeta} \leq 1 + 2\zeta$ for every $0 \leq \zeta \leq \frac{1}{2}$, and ensuring that $\frac{2nT\delta e^\varepsilon}{1-e^{-\varepsilon}} \leq \frac{1}{2}$, we get

$$
\begin{align}
(11) \quad &\leq \beta + \frac{2nT\delta e^\varepsilon}{1-e^{-\varepsilon}} + \left(1 + \frac{4nT\delta e^\varepsilon}{1-e^{-\varepsilon}}\right) \cdot \Pr[\mathcal{M}(\mathcal{A}_1(x_1),\ldots,\mathcal{A}_n(x_n)) \in F] \\
&\leq \beta + \frac{6nT\delta e^\varepsilon}{1-e^{-\varepsilon}} + \Pr[\mathcal{M}(\mathcal{A}_1(x_1),\ldots,\mathcal{A}_n(x_n)) \in F]
\end{align}
$$

This shows that `GenProt` increases the probability of any event by at most $\beta + \frac{6nT\delta e^\varepsilon}{1-e^{-\varepsilon}}$.
□

## 7 A Lower Bound via Anti-Concentration

In Section 3 we presented a LDP heavy-hitters algorithm with error $O\left(\frac{1}{\varepsilon}\sqrt{n \cdot \log\left(\frac{|X|}{\beta}\right)}\right)$. Ignoring the dependence on $\beta$, this error is known to be optimal. Specifically, Bassily and Smith [4] proved the following lower bound:

**Theorem 7.1** ([4]). *Let $\varepsilon = O(1)$ and $\delta = o\left(\frac{1}{n\log n}\right)$. Every non-interactive $(\varepsilon,\delta)$-LDP protocol for estimating the frequencies of elements from a domain $X$ must have worst-case error $\Omega\left(\frac{1}{\varepsilon}\sqrt{n \cdot \log |X|}\right)$ with constant probability.*

In this section we incorporate a tight dependence on the failure probability $\beta$ into the lower bound. Specifically, we show

**Theorem 7.2.** *Let $\varepsilon = O(1)$ and $\delta = o\left(\frac{1}{n\log n}\right)$. Every non-interactive $(\varepsilon,\delta)$-LDP protocol for estimating the frequencies of elements from a domain $X$ achieving worst-case error $\Delta$ with probability at least $1-\beta$ must have*

$$\Delta \geq \Omega\left(\frac{1}{\varepsilon}\sqrt{n \cdot \log\left(\frac{|X|}{\beta}\right)}\right).$$

In light of Theorem 7.1, we can fix $X = \{0,1\}$ and only show a lower bound of $\Omega\left(\frac{1}{\varepsilon}\sqrt{n \cdot \log(\frac{1}{\beta})}\right)$. To that end, we now strengthen an argument of Chan et al. [6] (see also [17, 34]) who obtained a lower bound of $\Omega(\sqrt{n})$ on the error:

**Theorem 7.3** ([6, 17, 34]). *Let $\varepsilon \leq 1/2$ and $\delta < 1$. Every $(\varepsilon, \delta)$-LDP frequency oracle must have worst-case error $\Omega(\sqrt{n})$ with constant probability.*

Before presenting the proof of Theorem 7.2, we establish some notation, and present a high-level overview of the proof. In this overview, we argue how the result follows from the ideas presented in the previous sections; namely, from advanced grouposition and from our generic transformation from approximate to pure LDP. However, we will present the complete proof in a somewhat more streamlined way that does not go through these generic results.

By our transformation from approximate to pure LDP, we may assume without loss of generality that we have a pure $\varepsilon$-LDP protocol $\mathcal{A}$ for $n$ users, which for every (distributed) database $D \in \{0,1\}^n$ counts the number of 1's in $D$ to within error $\Delta$ with success probability $1 - \beta$. Our goal is to show that $\Delta = \Omega\left(\frac{1}{\varepsilon}\sqrt{n \cdot \log(\frac{1}{\beta})}\right)$.

We generate an input to $\mathcal{A}$ as follows: For some constant $C > 0$, denote $m = C\varepsilon^2 n$, and let $S = (X_1, \ldots, X_m) \in \{0,1\}^m$ be a database chosen uniformly at random. Now define $D = (Y_1, \ldots, Y_n) \in \{0,1\}^n$, where $Y_i = X_{\lceil im/n \rceil}$. That is, the first $\frac{n}{m}$ elements in $D$ are $X_1$, the next $\frac{n}{m}$ elements are $X_2$, and so on. Now consider applying the protocol $\mathcal{A}$ onto the database $D$. The remainder of the proof proceeds as follows:

(i) On one hand, the error of $\mathcal{A}$ on $D$ is at most $\Delta$, and hence, renormalizing to $S$, we obtain an estimate of the number of 1's in $S$ with error at most $\frac{m}{n}\Delta = C\varepsilon^2 \Delta$.

(ii) On the other hand, every bit in $S$ has $\frac{n}{m} = \frac{1}{C\varepsilon^2}$ copies in $D$. By advanced grouposition, we provide every element in $S$ with say $(\frac{1}{100}, \frac{1}{100})$-DP. Using bounds on the mutual information of a random input and the output of a DP algorithm [4], we get that (a lot of) the elements in $S$ remain approximately uniform even conditioned on the transcript of the protocol (and they also remain independent).

Combining (i) and (ii), and applying an anti-concentration bound to the sum of elements in $S$, we will get that

$$C\varepsilon^2 \Delta \gtrsim \sqrt{|S| \cdot \log\left(\frac{1}{\beta}\right)} = \sqrt{C\varepsilon^2 n \cdot \log\left(\frac{1}{\beta}\right)},$$

which gives the result. We now present a full proof Theorem 7.2 which is based on these ideas, but does not need to go through our generic transformations.

*Proof of Theorem 7.2.* Assume that we have an $(\varepsilon, \delta)$-LDP protocol $\mathcal{A}$ for $n$ users, that for every (distributed) database $D \in \{0,1\}^n$ counts the number of 1's in $D$ to within error $\Delta$ with success probability $1 - \beta$. We generate an input to $\mathcal{A}$ as follows: For some constant $C > 0$, denote $m = C\varepsilon^2 n$, and let $S = (X_1, \ldots, X_m) \in \{0,1\}^m$ be a database chosen uniformly at random. Now define $D = (Y_1, \ldots, Y_n) \in \{0,1\}^n$, where $Y_i = X_{\lceil im/n \rceil}$. That is, the first $\frac{n}{m}$ elements in $D$ are $X_1$, the next $\frac{n}{m}$ elements are $X_2$, and so on.

Consider the execution of protocol $\mathcal{A}$ on the database $D = (Y_1, \ldots, Y_n)$, and let $R$ denote public randomness generated by the server. For $i \in [n]$, let $\mathcal{A}_i(R, \cdot)$ denote the $(\varepsilon, \delta)$-differentially private algorithm which acts on the bit $Y_i$, and let $A_i$ be a random variable denoting the outcome of $\mathcal{A}_i(R, Y_i)$. With this notation, we have that the outcome of the protocol is a deterministic function of $T = (R, A_1, \ldots, A_n)$, which we call the *transcript* of the protocol. For each fixed choice of $T = t$ we denote the outcome of the protocol as $\text{Est}_D(t)$. We also denote by $\text{Est}_S(t) = \frac{m}{n} \text{Est}_D(t)$ the renormalized estimate of the number of 1's in $S$. Observe that for every fixing of $X_1 = x_1, \ldots, X_m = x_m$ and $Y_1 = y_1, \ldots, Y_n = y_n$ we have

$$\left| \text{Est}_D(t) - \sum_{i=1}^{n} y_i \right| = \frac{m}{n} \cdot \left| \text{Est}_S(t) - \sum_{i=1}^{m} x_i \right|. \tag{12}$$

By construction, every bit $X_j \in S$ is used by the $\frac{n}{m}$ mechanisms

$$\mathcal{A}_{(j-1)\frac{n}{m}+1}, \mathcal{A}_{(j-1)\frac{n}{m}+2}, \ldots, \mathcal{A}_{j\frac{n}{m}}.$$

We abbreviate

$$\mathcal{B}_j(r, X_j) = (\mathcal{A}_{(j-1)\frac{n}{m}+1}(r, X_j), \ldots, \mathcal{A}_{j\frac{n}{m}}(r, X_j)).$$

That is, $\mathcal{B}_j(r, X_j)$ is the vector of outcomes of the $\frac{n}{m}$ mechanisms that operate on $X_j$ when the server's randomness is $R = r$. Let $C_2 > 1$ be a constant, and recall that $m = C\varepsilon^2 n$. By choosing $C = C(C_2)$ to be a large enough constant we have that $\mathcal{B}_j(r, \cdot)$ satisfies $(\frac{1}{C_2}, \frac{1}{C_2})$-differential privacy for every choice of $r$ (see advanced composition in [11]). We now use the following theorem to argue that, for every $r$, the mutual information between $X_j$ and $\mathcal{B}_j(r, X_j)$ is low:

**Theorem 7.4** ([4]). *Let $V$ be uniformly distributed over $\{0,1\}^d$. Let $\mathcal{Q} : \{0,1\}^d \to \mathcal{Z}$ be an $(\varepsilon, \delta)$-differentially private algorithm, and let $Z$ denote $\mathcal{Q}(V)$. Then, we have that*

$$I(V; Z) = O\left(\varepsilon^2 + \frac{\delta d}{\varepsilon} + \frac{\delta}{\varepsilon} \log(\varepsilon/\delta)\right).$$

Recall that $X_j$ is uniform in $\{0,1\}$, and that $\mathcal{B}_j(r, \cdot)$ satisfies $(\frac{1}{C_2}, \frac{1}{C_2})$-differential privacy. Hence, by choosing $C_2$ to be a large enough constant, for every $j \in [m]$ we have that $I(X_j; \mathcal{B}_j(r, X_j)) \leq \frac{1}{10}$, so

$$H(X_j | \mathcal{B}_j(r, X_j)) = H(X_j) - I(X_j; \mathcal{B}_j(r, X_j)) \geq \frac{9}{10}.$$

Recall that $H(X_j | \mathcal{B}_j(r, X_j))$ is defined as the average of $H(X_j | \mathcal{B}_j(r, X_j) = b_j)$ over all possible values $b_j$ in the range of $\mathcal{B}_j(r, \cdot)$. So,

$$\frac{9}{10} \leq \sum_{b_j} \Pr[\mathcal{B}_j(r, X_j) = b_j] \cdot H(X_j | \mathcal{B}_j(r, X_j) = b_j)$$

$$\leq \sum_{b_j : H(X_j | \mathcal{B}_j(r, X_j) = b_j) \geq \frac{1}{2}} \Pr[\mathcal{B}_j(r, X_j) = b_j] + \sum_{b_j : H(X_j | \mathcal{B}_j(r, X_j) = b_j) < \frac{1}{2}} \Pr[\mathcal{B}_j(r, X_j) = b_j] \cdot \frac{1}{2}$$

$$= \Pr\left[\mathcal{B}_j(r, X_j) = b_j \text{ s.t. } H(X_j | \mathcal{B}_j(r, X_j) = b_j) \geq \frac{1}{2}\right]$$

$$+ \frac{1}{2} \cdot \Pr\left[\mathcal{B}_j(r, X_j) = b_j \text{ s.t. } H(X_j | \mathcal{B}_j(r, X_j) = b_j) < \frac{1}{2}\right]$$

$$\leq \Pr\left[\mathcal{B}_j(r, X_j) = b_j \text{ s.t. } H(X_j | \mathcal{B}_j(r, X_j) = b_j) \geq \frac{1}{2}\right] + \frac{1}{2}.$$

Let $t = (r, a_1, \ldots, a_n)$ be a possible transcript, and let $b_j$ denote the portion of $t$ corresponding to $\mathcal{B}_j$, that is, $b_j = (a_{(j-1)\frac{n}{m}+1}, \ldots, a_{j\frac{n}{m}})$. We say that index $j \in [m]$ is *good* for the transcript $t$ if

$$H(X_j | \mathcal{B}_j(r, X_j) = b_j) \geq 1/2.$$

Then, for every $r$ and every $j \in [m]$, we have that

$$\Pr[T = t \text{ s.t. } j \text{ is good for } t | R = r] = \Pr\left[\mathcal{B}_j(r, X_j) = b_j \text{ s.t. } H(X_j | \mathcal{B}_j(r, X_j) = b_j) \geq \frac{1}{2}\right] \geq \frac{2}{5},$$

where the second probability is only over $X_j$ and the coins of $\mathcal{B}_j(r, X_j)$. Consider the following event:

> **Event $E_1$ (over sampling $S \in \{0,1\}^m$ and the execution of $\mathcal{A}$):**
> $T = t$ s.t. there are at least $\frac{m}{5}$ choices for $j \in [m]$ which are good for $t$.

For every fixture of $R = r$, by the Chernoff bound, assuming that $m \geq 20 \ln(3/\beta)$, we have that $\Pr[E_1 | R = r] \geq 1 - \frac{\beta}{3}$. Hence, $\Pr[E_1] \geq 1 - \frac{\beta}{3}$. We continue with the proof assuming that event $E_1$ occurs.

Fix a transcript $t = (r, a_1, \ldots, a_n)$, and let $j$ be a good index for $t$. We have that $b_j$ is s.t. $H(X_j | \mathcal{B}_j(r, X_j) = b_j) \geq 1/2$. Hence,

$$\frac{1}{10} \leq \Pr[X_j = 1 | \mathcal{B}_j(r, X_j) = b_j] \leq \frac{9}{10}.$$

Now observe that the executions of $B_j(r, \cdot)$ are independent across different $j$'s, and are independent of the server's randomness $R$ (since we fixed $r$ in $B_j(r, \cdot)$). Hence

$$X_j | \mathcal{B}_j(r, X_j) = b_j \equiv X_j | T = t.$$

So we also have that $\frac{1}{10} \leq \Pr[X_j = 1 | T = t] \leq \frac{9}{10}$, and hence,

$$\text{Var}[X_j | T = t] \geq 9/100.$$

Moreover, for every choice of $t$ it holds that the random variables $(X_1 | T=t), (X_2 | T=t), \ldots, (X_m | T=t)$ are independent (this is a general fact about interactive protocols – if the parties' inputs start independent, they remain independent conditioned on the transcript).

Fix a transcript $t$. As event $E_1$ has occurred, there is a set $G \subseteq [m]$ of size $\frac{m}{5}$ of good indexes for $t$. We now use the following anti-concentration theorem to argue about $\left(\sum_{i \in M} X_i \mid T=t\right)$.

**Theorem 7.5.** *[13, 23] There exist constants $a, b > 0$ s.t. the following holds. Let $X$ be a sum of independent random variables, each attaining values in $[0, 1]$, and let $\sigma = \sqrt{\text{Var}[X]} \geq 200$. Then for all $t \in \left[0, \frac{\sigma^2}{100}\right]$, we have*

$$\Pr[X \geq \mathbb{E}[X] + t] \geq a \cdot e^{-bt^2/\sigma^2}.$$

An immediate corollary of this theorem is following:

**Corollary 7.6.** *There exist constants $a, b, c > 0$ s.t. the following holds. Let $X$ be a sum of independent random variables, each attaining values in $[0, 1]$, and let $\sigma = \sqrt{\text{Var}[X]} \geq 200$. Then for every $a \geq \beta \geq 2^{-b\sigma^2}$ and every interval $I \in \mathbb{R}$ of length $|I| \leq c \cdot \sqrt{\sigma^2 \cdot \log(\frac{a}{\beta})}$ we have*

$$\Pr[X \notin I] \geq \beta.$$

For completeness, we give in Appendix A a simple proof of a weaker version of corollary that suffices for our application. Applying Corollary 7.6 to the random variables $(X_j|T{=}t)$ for $j \in G$, we get that there are constants $a, b, c > 0$ s.t. for every $a \geq \beta \geq 2^{-b\varepsilon^2 n}$ and every interval $I \in \mathbb{R}$ of length $|I| \leq c \cdot \sqrt{\varepsilon^2 n \cdot \log(\frac{a}{\beta})}$ it holds that

$$\Pr\left[\sum_{i \in G} X_i \notin I \,\bigg|\, T{=}t\right] \geq 2\beta. \tag{13}$$

Recall that, even conditioned on $T{=}t$, the random variables $X_1, \ldots, X_m$ are independent. Hence, Inequality (13) also holds for the sum of all $X_j$ (since Inequality 13 holds for every fixed choice of $\{X_j : j \notin G\}$). That is,

$$\Pr\left[\sum_{i \in [m]} X_i \notin I \,\bigg|\, T{=}t\right] \geq 2\beta.$$

In particular, consider the interval $I(t) = \text{Est}_S(t) \pm \frac{c}{2} \cdot \sqrt{\varepsilon^2 n \cdot \log\left(\frac{a}{\beta}\right)}$. We have that

$$\begin{aligned}
\Pr\left[\left|\text{Est}_S(T) - \sum_{i \in [m]} X_i\right| > \frac{c}{2} \cdot \sqrt{\varepsilon^2 n \cdot \log\left(\frac{a}{\beta}\right)}\right] &= \Pr\left[\sum_{i \in [m]} X_i \notin I(T)\right] \\
&\geq \Pr[E_1] \cdot \Pr\left[\sum_{i \in [m]} X_i \notin I(T) \,\bigg|\, E_1\right] \\
&\geq \left(1 - \frac{\beta}{3}\right) \cdot 2\beta > \beta.
\end{aligned}$$

And finally, using Equation (12),

$$\Pr\left[\left|\text{Est}_D(T) - \sum_{i \in [n]} Y_i\right| > \frac{c}{2C\varepsilon} \cdot \sqrt{n \cdot \log\left(\frac{a}{\beta}\right)}\right] > \beta.$$

That is, with probability greater than $\beta$ the protocol has error at least $\Omega\left(\frac{1}{\varepsilon}\sqrt{n \cdot \log\left(\frac{1}{\beta}\right)}\right)$ when estimating the number of ones in the random database $D \in \{0,1\}^n$. Hence, there must exist some fixed database on which it has error $\Omega\left(\frac{1}{\varepsilon}\sqrt{n \cdot \log\left(\frac{1}{\beta}\right)}\right)$ with probability greater than $\beta$. $\square$

**Acknowledgements.** We thank Kobbi Nissim for suggesting the name "advanced grouposition" for the phenomenon described in Section 4.

# A  Simple Variant of Corollary 7.6

In Section 7 we used the following anti-concentration result:

**Corollary 7.6** ([13, 23])**.** *There exist constants $a, b, c > 0$ s.t. the following holds. Let $X$ be a sum of independent random variables, each attaining values in $[0, 1]$, and let $\sigma = \sqrt{\mathrm{Var}[X]} \geq 200$. Then for every $a \geq \beta \geq 2^{-b\sigma^2}$ and every interval $I \in \mathbb{R}$ of length $|I| \leq c \cdot \sqrt{\sigma^2 \cdot \log(\frac{a}{\beta})}$ we have*

$$\Pr[X \notin I] \geq \beta.$$

This result has relatively simple proofs in the special case where $X$ is the sum of *i.i.d.* random variables (see, e.g., [21]). However, we were unable to find a reference with a simple analysis for the case where the variables are independent but not necessarily identically distributed. In particular, the results of Feller [13] are quite involved, as they are much more general than Corollary 7.6. In this section we provide a simple analysis for a special case of Corollary 7.6 that suffices for our applications in Section 7.

Let $X_1, \ldots, X_n$ be independent random variables, each attaining values in $\{0, 1\}$, and denote $p_i = \Pr[X_i = 1]$. We assume that, for some constant $c \in (0, \frac{1}{2})$, for all $i \in [n]$ we have that $\frac{1}{2} - c \leq p_i \leq \frac{1}{2} + c$.

Our goal is to show that for any interval $I$ of length $|I| \leq O\left(\sqrt{n \cdot \log(\frac{1}{\beta})}\right)$ it holds that $\Pr[\sum_i X_i \notin I] \geq \beta$. The first step is to relate the situation to the easier-to-analyze case of *i.i.d.* random variables, using the following lemma.

**Lemma A.1.** *Let $X, Y$ be independent random variables, where $Y \in \mathbb{R}$ and $X \in \{0, 1\}$ with $\frac{1}{2} - c \leq p = \mathbb{E}[X] \leq \frac{1}{2} + c$. Also let $X^+, X^- \in \{0, 1\}$ be independent of $Y$ s.t. $p^+ = \mathbb{E}[X^+] = \frac{1}{2} + c$ and $p^- = \mathbb{E}[X^-] = \frac{1}{2} - c$. Then for every interval $I \subseteq \mathbb{R}$ we have,*

$$\Pr[X + Y \notin I] \geq \min\left\{\Pr[X^+ + Y \notin I], \Pr[X^- + Y \notin I]\right\}.$$

*Proof.*

$$\Pr[X + Y \notin I] = \sum_{y \in \mathbb{R}} \Pr_Y[Y = y] \cdot \Pr_X[X + y \notin I]$$

$$= \Pr_Y \begin{bmatrix} Y + 1 \notin I \\ \text{and} \\ Y \notin I \end{bmatrix} \cdot 1 + \underbrace{\Pr_Y \begin{bmatrix} Y + 1 \notin I \\ \text{and} \\ Y \in I \end{bmatrix}}_{\text{Expression } A} \cdot p + \underbrace{\Pr_Y \begin{bmatrix} Y + 1 \in I \\ \text{and} \\ Y \notin I \end{bmatrix}}_{\text{Expression } B} \cdot (1 - p).$$

(14)

Trivially, one of these two expressions is greater or equal to the other. Let us define $\tilde{p}$ to be $\tilde{p} = \frac{1}{2} - c$ if $A \geq B$, and $\tilde{p} = \frac{1}{2} + c$ otherwise. Furthermore, let $\tilde{X} \in \{0, 1\}$ be s.t. $\mathbb{E}[\tilde{X}] = \tilde{p}$. With this notation we have that

$$(14) \geq \Pr_Y \begin{bmatrix} Y+1 \notin I \\ \text{and} \\ Y \notin I \end{bmatrix} \cdot 1 + \Pr_Y \begin{bmatrix} Y+1 \notin I \\ \text{and} \\ Y \in I \end{bmatrix} \cdot \tilde{p} + \Pr_Y \begin{bmatrix} Y+1 \in I \\ \text{and} \\ Y \notin I \end{bmatrix} \cdot (1 - \tilde{p})$$

$$= \Pr[\tilde{X} + Y \notin I] \geq \min\left\{\Pr[X^+ + Y \notin I], \Pr[X^- + Y \notin I]\right\}.$$

□

Consider again our independent random variables $X_1, \ldots, X_n \in \{0, 1\}$, and let $I \subseteq \mathbb{R}$ be an interval. Using Lemma A.1 we can replace, one-by-one, every variable $X_i$ with a variable $\tilde{X}_i$ with expectation either $\frac{1}{2} + c$ or $\frac{1}{2} - c$ (exactly), without decreasing the probability that $\sum_i X_i \notin I$. That is, we have established the following statement.

**Corollary A.2.** *Let $X_1, \ldots, X_n \in \{0, 1\}$ be independent, where $\frac{1}{2} - c \leq p_i = \mathbb{E}[X_i] \leq \frac{1}{2} + c$. For every interval $I \subseteq \mathbb{R}$ there exist a collection of $n$ independent random variables $\tilde{X}_1, \ldots, \tilde{X}_n \in \{0, 1\}$ where $\mathbb{E}[\tilde{X}_i] = \tilde{p}_i \in \{\frac{1}{2} + c, \frac{1}{2} - c\}$, such that*

$$\Pr\left[\sum_i X_i \notin I\right] \geq \Pr\left[\sum_i \tilde{X}_i \notin I\right].$$

So, we have a collection of $n$ random variables $\tilde{X}_1, \ldots, \tilde{X}_n$ of two possible types (either $\tilde{p}_i = \frac{1}{2} + c$ or $\tilde{p}_i = \frac{1}{2} - c$). Clearly, at least one type appears at least $n/2$ times. We hence let $G \subseteq [n]$ denote a set of size $n/2$ s.t. for all $i \in G$ we have that $\tilde{X}_i$ are of the same type. We can now apply again Lemma A.1 to our random variables $\tilde{X}_1, \ldots, \tilde{X}_n$, and for every $i \notin G$ replace $\tilde{X}_i$ with a constant 0 or 1 (this amounts to shifting the interval $I$). This brings us to the following corollary:

**Corollary A.3.** *Let $X_1, \ldots, X_n \in \{0, 1\}$ be independent, where $\frac{1}{2} - c \leq p_i = \mathbb{E}[X_i] \leq \frac{1}{2} + c$. For every interval $I \subseteq \mathbb{R}$ there exist an interval $\hat{I} \subseteq \mathbb{R}$ of the same length $|\hat{I}| = |I|$ and $\hat{p} \in \{\frac{1}{2} + c, \frac{1}{2} - c\}$ s.t. the following holds.*

$$\Pr\left[\sum_i X_i \notin I\right] \geq \Pr\left[\text{Bin}\left(\frac{n}{2}, \hat{p}\right) \notin \hat{I}\right].$$

The result now follows from anti-concentration bounds for the Binomial distribution. For instance, we can apply the following bound.

**Theorem A.4** ([21, Lemma 5.2]). *Let $0 < p \leq \frac{1}{2}$, and let $\sqrt{3np} \leq t \leq np/2$. Then,*

$$\Pr\left[\text{Bin}(n, p) \leq np - t\right] \geq \exp\left(-\frac{9t^2}{np}\right),$$

$$\Pr\left[\text{Bin}(n, p) \geq np + t\right] \geq \exp\left(-\frac{9t^2}{np}\right).$$

To complete the proof, let $I \subseteq \mathbb{R}$ be an interval of length at most $t$. Let $\hat{I} = [\hat{a}, \hat{b}]$ and $\hat{p}$ be as in Corollary A.3. We have that

$$\Pr\left[\text{Bin}\left(\frac{n}{2}, \hat{p}\right) \notin \hat{I}\right] = \Pr\left[\text{Bin}\left(\frac{n}{2}, \hat{p}\right) < \hat{a}\right] + \Pr\left[\text{Bin}\left(\frac{n}{2}, \hat{p}\right) > \hat{b}\right]. \quad (15)$$

There are two cases: Either $\hat{p} = \frac{1}{2} - c$ or $\hat{p} = \frac{1}{2} + c$. As the two cases are symmetric, we now proceed assuming that $\hat{p} = \frac{1}{2} - c$.

Recall that the median of $\text{Bin}\left(\frac{n}{2}, \hat{p}\right)$ is between $\lfloor \frac{n\hat{p}}{2} \rfloor$ and $\lceil \frac{n\hat{p}}{2} \rceil$. Hence, if $\hat{b} \leq \frac{n\hat{p}}{2} - 1$, then $\Pr\left[\text{Bin}\left(\frac{n}{2}, \hat{p}\right) > \hat{b}\right] \geq \frac{1}{2}$ and the proof is complete. We therefore proceed assuming that $\hat{b} \geq \frac{n\hat{p}}{2} - 1$, and hence, $\hat{a} \geq \frac{n\hat{p}}{2} - 1 - t \geq \frac{n\hat{p}}{2} - 2t$. So,

$$(15) \geq \Pr\left[\text{Bin}\left(\frac{n}{2}, \hat{p}\right) < \frac{n\hat{p}}{2} - 2t\right] \geq \exp\left(-\frac{72t^2}{\hat{p}n}\right),$$

where the last inequality follows from Theorem A.4 by asserting that $\sqrt{3\hat{p}n} \leq 3t \leq \hat{p}n/4$. This results in the following theorem:

**Theorem A.5.** *There exist constants $a, b, c > 0$ s.t. the following holds. Let $X_1, \ldots, X_n \in \{0, 1\}$ be independent random variables, such that for every $i \in [n]$ we have $\frac{1}{10} \leq \mathbb{E}[X_i] \leq \frac{9}{10}$. Then for every $a \geq \beta \geq 2^{-bn}$ and every interval $I \in \mathbb{R}$ of length $|I| \leq c \cdot \sqrt{n \cdot \log(\frac{1}{\beta})}$ we have*

$$\Pr[X \notin I] \geq \beta.$$

# B  Proof of Theorem 3.6 [22]

Recall the definition of unique-list-recoverable codes:

**Definition 3.5.** *An $(\alpha, \ell, L)$-unique-list-recoverable code is a pair of mappings $(\text{Enc}, \text{Dec})$ where $\text{Enc} : X \to ([Y] \times [Z])^M$, and $\text{Dec} : (([Y] \times [Z])^\ell)^M \to X^L$, such that the following holds. Let $L_1, \ldots, L_M \in ([Y] \times [Z])^\ell$. Assume that for every $m \in [M]$, if $(y, z), (y', z') \in L_m$ then $y \neq y'$. Then for every $x \in X$ satisfying $|\{m : \text{Enc}(x)_m \in L_m\}| \geq (1 - \alpha)M$ we have that $x \in \text{Dec}(L_1, \ldots, L_M)$.*

The following theorem is a direct consequence of the results of [22] (appearing implicitly in their analysis). We include their proof here for completeness.

**Theorem 3.6** ([22]). *There exist constants $C > 1$ and $0 < \alpha < 1$ such that the following holds. For all constants $M \leq \log |X|$ and $Y, \ell \in \mathbb{N}$, and for every fixed choice of functions $h_1, \ldots, h_M : X \to [Y]$, there is a construction of an $(\alpha, \ell, L)$-unique-list-recoverable code*

$$\text{Enc} : X \to ([Y] \times [Z])^M \qquad \text{and} \qquad \text{Dec} : (([Y] \times [Z])^\ell)^M \to X^L,$$

*where $L \leq C \cdot \ell$ and $Z \leq (|X|^{1/M} \cdot Y)^C$. Furthermore, there is a mapping $\widetilde{\text{Enc}} : X \to [Z]^M$ such that for every $x \in X$ we have $\text{Enc}(x) = \left((h_1(x), \widetilde{\text{Enc}}(x)_1), \ldots, (h_M(x), \widetilde{\text{Enc}}(x)_M)\right).$*

We use the following tools in the construction:

1. We will use a (standard) error-correcting code (enc, dec) with constant rate that can correct an $\Omega(1)$-fraction of errors. Such codes exist with linear time encoding and decoding [32, 15]. We partition $\text{enc}(x)$ into $M$ contiguous bitstrings of equal length $O(\log |X|)/M$, and let $\text{enc}(x)_m$ denote the $m^{\text{th}}$ bitstring (for $m \in [M]$). Note that we write (enc, dec) to denote this (standard) error correction code, and (Enc, Dec) for the unique-list-recoverable code that we are constructing.

2. We also use a $d$-regular $\lambda_0$-spectral expander[6] $F$ on $M$ vertices for some $d = O(1)$, where $\lambda_0 = \alpha d$ for some (small) constant $\alpha > 0$ to be specified later. Such an $F$ can be constructed in time $\text{poly}(M)$ with $d = \text{poly}(1/\alpha)$ deterministically, for every $M$ of the form $M = D^i$ for some constant $D$. See [28].[7]

We now define $\widetilde{\text{Enc}} : X \to [Z]^M$ and $\text{Enc} : X \to ([Y] \times [Z])^M$. For $m \in [M]$ we define:

$$\widetilde{\text{Enc}}(x)_m = \left(\text{enc}(x)_m, h_{\Gamma(m)_1}(x), \ldots, h_{\Gamma(m)_d}(x)\right) \quad \text{and} \quad \text{Enc}(x)_m = \left(h_m(x), \widetilde{\text{Enc}}(x)_m\right),$$

where $\Gamma(m)_k$ is the $k^{\text{th}}$ neighbor of $m$ in $F$. Observe that $\log Z = O(\log |X|)/M + d \cdot \log Y$, and hence, $Z \leq (|X|^{1/M} \cdot Y)^C$ for some constant $C$.

We now explain how the decoder Dec operates. To that end, let $L_1, \ldots, L_M \in ([Y] \times [Z])^\ell$ be such that for every $m \in [M]$, if $(y, z), (y', z') \in L_m$ then $y \neq y'$.

We construct the following graph $G$ on the layered vertex set $V = [M] \times [Y]$. For $m \in [M]$, we can view each element $(y, z) \in L_m$ as suggesting $d$ edges to add to $G$. Specifically, if $z = (e, y_1, \ldots, y_d)$, then this suggests connecting $(m, y)$ with each of $(\Gamma(m)_1, y_1), \ldots, (\Gamma(m)_d, y_d)$. We then let $G$ be the graph created by including the at most $(d/2) \cdot \sum_m |L_m|$ edges suggested by the elements across all $L_m$'s (we only include an edge if both endpoints suggest it).

Fix a domain element $x \in X$ satisfying $|\{m : \text{Enc}(x)_m \in L_m\}| \geq (1 - \alpha)M$. The goal of the decoder Dec is to recover $x$. Let $W$ be the set of $M$ vertices $\{(m, h_m(x))\}_{m=1}^M$. Consider first the ideal case in which the encodings $\text{Enc}(x)_m$ appear throughout *all* of the lists $L_m$, i.e., $|\{m : \text{Enc}(x)_m \in L_m\}| = M$. In this case, $W$ would be an isolated connected component in $G$, and furthermore the induced graph on $W$ would be $F$. We will use the following lemma (a version of this lemma is known as the expander mixing lemma):

**Lemma B.1** ([1]). *Let $A$ be the adjacency matrix of a $d$-regular graph with vertex set $V$. Suppose the second largest eigenvalue of $A$ in magnitude is $\lambda > 0$. Then for any $S \subseteq V$, writing $|S| = r|V|$, $|\partial S| \geq (d - \lambda)(1 - r)|S|$.*

By Lemma B.1, in the ideal case, for any subset $A$ of $W$ with $|A| = r|W| = rM$ we have

$$|E(A, W \setminus A)| \geq (d - \lambda_0)(1 - r)rM \geq \left(r(1 - r) - \frac{\lambda_0}{4d}\right) dM.$$

Let us now understand the possible effects of "bad" indexes on $W$, where we say that $m \in [M]$ is bad if $\text{Enc}(x)_m \notin L_m$. We can no longer argue that $W$ is an isolated copy of $F$, but what we can say, even with $\alpha M$ bad indexes, is that $W$ forms an $O(\alpha)$-spectral cluster:

---

[6]We say that a graph is a $\lambda$-spectral expander if the second largest eigenvalue, in magnitude, of its unnormalized adjacency matrix is at most $\lambda$.

[7]The assumption that $M = D^i$ can be removed as follows. The construction only needs a spectral expander, and not an edge expander, and spectral expansion can be verified efficiently. As a random graph is a spectral expander with high probability, so we can construct an expander for every $M$ in efficient Las Vegas time.

**Definition B.2.** *An $\eta$-spectral cluster is a vertex set $W \subseteq V$ of any size satisfying the following two conditions: First, only an $\eta$-fraction of the edges incident to $W$ leave $W$, that is, $|\partial(W)| \leq \eta \cdot vol(W)$, where $vol(W)$ is the sum of edge degrees of vertices inside $W$. Second, given any subset $A$ of $W$, let $r = vol(A)/vol(W)$ and $B = W \setminus A$. Then*

$$|E(A,B)| \geq (r(1-r) - \eta)vol(W).$$

*Note $r(1-r)vol(W)$ is the number of edges one would expect to see between $A$ and $B$ had $W$ been a random graph with a prescribed degree distribution.*

Suppose $m$ is bad. First, $W$ might lose at most $d$ edges from $F$ (those corresponding to edges incident upon vertex $m$ in $F$). Second, $L_m$ might contain an element $(y, z')$ s.t. $y = h_m(x)$, but $z' \neq \widetilde{\mathrm{Enc}}(x)$, possibly causing edges to be included in $W$ which do not appear in $F$, or even inserting edges that cross the cut $(W, G \setminus W)$. At most $d$ such edges are inserted in this way. Thus, across all bad indexes, the total number of $F$-edges deleted within $W$ is at most $\alpha dM$, and the total number of edges crossing the cut $(W, G \setminus W)$ is also at most $\alpha dM$. Also, the volume $vol(W)$ is always at most $dM$ and at least $(1-\alpha)dM$. Thus after considering bad indexes, for any subset $A$ of $W$ as above,

$$|E(A, W \setminus A)| \geq \left(r(1-r) - \frac{\lambda_0}{4d} - \alpha\right)dM \geq (r(1-r) - \alpha_0)vol(W).$$

for $\alpha_0 \geq \alpha + \lambda_0/(4d) = 5\alpha/4$. Furthermore, the number of edges leaving $W$ to $G \setminus W$ is

$$|\partial(W)| \leq \alpha dm \leq \frac{\alpha}{1-\alpha}vol(W) \leq \frac{5\alpha}{4}vol(W) \leq \alpha_0 vol(W),$$

for $\alpha \leq 1/4$. Thus $W$ is an $\alpha_0$-spectral cluster in $G$ representing $x$. Our task of identifying $W$ therefore reduces to a clustering problem (identifying spectral clusters in $G$), which we can solve using the following algorithm:

**Theorem B.3** ([22]). *There exists universal constants $\eta_0$ and $K$ such the following holds. There is a polynomial time and linear space algorithm $\mathcal{A}$ that, for any given $\eta \leq \eta_0$ and graph $G = (V, E)$, finds a family of disjoint vertex sets $U_1, \ldots U_\ell$ so that every $\eta$-spectral cluster $U^*$ of $G$ matches some set $U_i$ in the sense that*

- $vol(U^* \setminus U_i) \leq 3\eta \cdot vol(U^*)$.
- $vol(U_i \setminus U^*) \leq K\eta \cdot vol(U^*)$.

Applying the above theorem to our graph $G$, we have that algorithm $\mathcal{A}$ recovers a $W'$ missing at most $3\alpha_0 vol(W)$ volume from $W$, and containing at most $K\alpha_0 vol(W)$ volume from $G \setminus W$.

After obtaining $W'$, we remove any vertex from $W'$ of degree $\leq d/2$, so $W'$ contains at most $2K\alpha_0 vol(W)/d \leq 2K\alpha_0 M$ vertices from outside $W$. Furthermore, since there are at most $\alpha dM$ edges lost from $W$ due to bad indexes, at most $2\alpha M$ vertices in $W$ had their degrees reduced to $\leq d/2$, and thus removing low-degree vertices removed at most $2\alpha M$ vertices from $W$. Also, as $W'$ is missing at most $3\alpha_0 vol(W)$ volume from $W$, we have that $W'$ misses at most $6\alpha_0 M$ vertices with degree $> d/2$ from $W$. Overall, $W'$ contains at most $2K\alpha_0 M$ vertices from outside $W$, and misses at most $(6\alpha_0 + 2\alpha)M$ vertices from $W$.

We then form a (corrupted) codeword $c'$ by concatenating encoding chunks specified by the vertices in $W'$. Since then at most a $((2K+6)\alpha_0 + 2\alpha)$-fraction of entries in $\text{enc}(x)$ and $c'$ differ, for $\alpha$ sufficiently small, we successfully decode and obtain the binary encoding of $x$.

It remains to show that the number of identified elements (denoted as $L$) is small. To that end, observe that in the analysis above we had $vol(W') \geq vol(W) - vol(W \setminus W') \geq (1 - \alpha - 3\alpha_0)dM$. Moreover, recall that the total number of edges in $G$ is at most $\ell dM$, and hence, there could be at most $\frac{2\ell dM}{(1-\alpha-3\alpha_0)dM} = O(\ell)$ such clusters $W'$. So $L = O(\ell)$.